\documentclass[lettersize,journal]{IEEEtran}

\IEEEoverridecommandlockouts
\ifCLASSOPTIONcompsoc
    \usepackage[nocompress]{cite}
\else
    \usepackage{cite}
\fi
\usepackage{cite}
\ifCLASSOPTIONcompsoc
    \usepackage[caption=false,font=footnotesize,labelfont=sf,textfont=sf]{subfig}
\else
    \usepackage[caption=false,font=footnotesize]{subfig}
\fi
\ifCLASSOPTIONcaptionsoff
    \usepackage[nomarkers]{endfloat}
\let\MYoriglatexcaption\caption
\renewcommand{\caption}[2][\relax]{\MYoriglatexcaption[#2]{#2}}
\fi

\usepackage[T1]{fontenc}
\usepackage{url}
\usepackage{multicol}
\usepackage{multirow}
\usepackage{subfig}
\usepackage{dsfont}
\usepackage{hyperref}
\usepackage{bm}
\usepackage[pdftex]{graphicx}
\usepackage{array,amsmath,amssymb,amsfonts}
\usepackage{amsthm}
\usepackage{subeqnarray}
\usepackage{cases}
\usepackage{booktabs}
\usepackage{color}
\usepackage[ruled,lined,linesnumbered,commentsnumbered,longend]{algorithm2e}
\makeatletter
\newcommand{\removelatexerror}{\let\@latex@error\@gobble}
\makeatother

\SetCommentSty{mycommfont}

\def\BibTeX{{\rm B\kern-.05em{\sc i\kern-.025em b}\kern-.08em
    T\kern-.1667em\lower.7ex\hbox{E}\kern-.125emX}}
\vspace{-0.3cm}
\IEEEoverridecommandlockouts
\usepackage{makecell}

\usepackage{listings}
\definecolor{dkgreen}{rgb}{0,0.6,0}
\definecolor{gray}{rgb}{0.5,0.5,0.5}
\definecolor{mauve}{rgb}{0.58,0,0.82}
\usepackage{siunitx} 
\begin{document}

\title{Cloud-Native Computing: A Survey from the Perspective of Services}
\author{
        Shuiguang~Deng,~\IEEEmembership{Senior~Member,~IEEE,} 
        Hailiang~Zhao,
        Binbin~Huang,
        Cheng~Zhang,
        Feiyi~Chen,
        Yinuo~Deng,
        Jianwei~Yin,
        Schahram~Dustdar,~\IEEEmembership{Fellow,~IEEE},
        and~Albert~Y.~Zomaya,~\IEEEmembership{Fellow,~IEEE}
  \thanks{S. Deng, H. Zhao, C. Zhang, F. Chen, Y. Deng, and J. Yin are with the College of Computer Science and Technology, Zhejiang University, Hangzhou 310027, China. E-mail: {dengsg, hliangzhao, coolzc, chenfeiyi, yinuo, zjuyjw}@zju.edu.cn}.
  \thanks{B. Huang is with the College of Computer Science and Technology, Hangzhou Dianzi University, Hangzhou 310012, China. E-mail: {huangbinbin@hdu.edu.cn}.}
  \thanks{S. Dustdar is with the Distributed Systems Group, Technische Universität Wien, 1040 Vienna, Austria. E-mail: dustdar@dsg.tuwien.ac.at.}
  \thanks{A. Y. Zomaya is with the School of Computer Science, University of Sydney, Sydney, NSW 2006, Australia. e-mail: albert.zomaya@sydney.edu.au.}
}

\IEEEtitleabstractindextext{%
\begin{abstract}
The development of cloud computing delivery models inspires the emergence of cloud-native computing. Cloud-native computing, as the most influential development principle for web applications, has already attracted increasingly more attention in both industry and academia. Despite the momentum in the cloud-native industrial community, a clear research roadmap on this topic is still missing. As a contribution to this knowledge, this paper surveys key issues during the life-cycle of cloud-native applications, from the perspective of services. Specifically, we elaborate the research domains by decoupling the life-cycle of cloud-native applications into four states: building, orchestration, operate, and maintenance. We also discuss the fundamental necessities and summarize the key performance metrics that play critical roles during the development and management of cloud-native applications. We highlight the key implications and limitations of existing works in each state. The challenges, future directions, and research opportunities are also discussed.
\end{abstract}

}

\maketitle

\IEEEdisplaynontitleabstractindextext

\ifCLASSOPTIONpeerreview
\begin{center} \bfseries EDICS Category: 3-BBND \end{center}
\fi
%
\IEEEpeerreviewmaketitle

\section{Introduction}\label{s1}
\IEEEPARstart{S}ervices are self-describing and technology-neutral computation entities that support rapid and low-cost composition of web applications in distributed network systems \cite{zhl-service-computing-survey}. Service-Oriented Architecture (SOA) is the principle to design the software systems by $(i)$ provisioning independent, reusable, and automated functions as reusable services and $(ii)$ providing a robust and secure foundation for leveraging these services \cite{zhl-service-computing-survey2}. In recent years, the most influential variant of SOA is \textit{the microservices architecture}, which decouples a monolithic application into a collection of loosely-coupled, fine-grained microservices, communicating through lightweight protocols \cite{zhl-microservice1}. Over the last decade, the microservices approach is more and more appealing, as it allows teams and software organizations to be more productive to build continuously deployed systems with the support of DevOps \cite{zhl-leite2019survey} and continuous integration/continuous delivery (CI/CD) pipelines \cite{zhl-zampetti2021ci}. 

Accompanied with the development of microservices, a new terminology, \textit{cloud-native}, or \textit{cloud-native computing}, is attracting increasingly more attention in academia. In accordance with Cloud Native Computing Foundation (CNCF), the open source, vendor-neutral hub of cloud-native computing\footnote{https://www.cncf.io/about/who-we-are/}, cloud-native is the collection of technologies that \textit{break down applications into microservices and package them in lightweight containers to be deployed and orchestrated across a variety of servers\footnote{https://github.com/cncf/toc/blob/main/DEFINITION.md}}. In addition to the microservices architecture, cloud-native is also characterized by the following terminologies:
\begin{itemize}
  \item \textit{Containerization}. Containerization is a function isolation mechanism which leverages the Linux kernel to isolate resources, and creating containers as different processes in Host OS \cite{zhl-barlev2016secure,zhl-mattetti2015securing}. Docker, with a ten-year development, is the most popular implementation of the containerization techniques \cite{zhl-Docker}. Combing containerization with the microservices architecture, each part of an application, including processes, libraries, etc., is packaged into its own container. This facilitates reproducibility, transparency, and resource isolation.
  \item \textit{Orchestration}. Orchestration is the automated configuration, management, and coordination of the inter-related microservices to build the elastic and scalable functionalities. Because microservices are deployed in the way of containers, orchestration reduces to the automation of the operational effort to manage the containers' life-cycle, including resource provisioning, deployment, scheduling, scaling (up and down), networking, load balancing, etc, in order to execute the applications' workflows or processes. Kubernetes \cite{zhl-Kubernetes}, originated from Google's Borg cluster manager \cite{zhl-verma2015large}, is the most popular open-source container orchestration software.
\end{itemize}

\begin{figure*}
  \centering
  \includegraphics[width=7in]{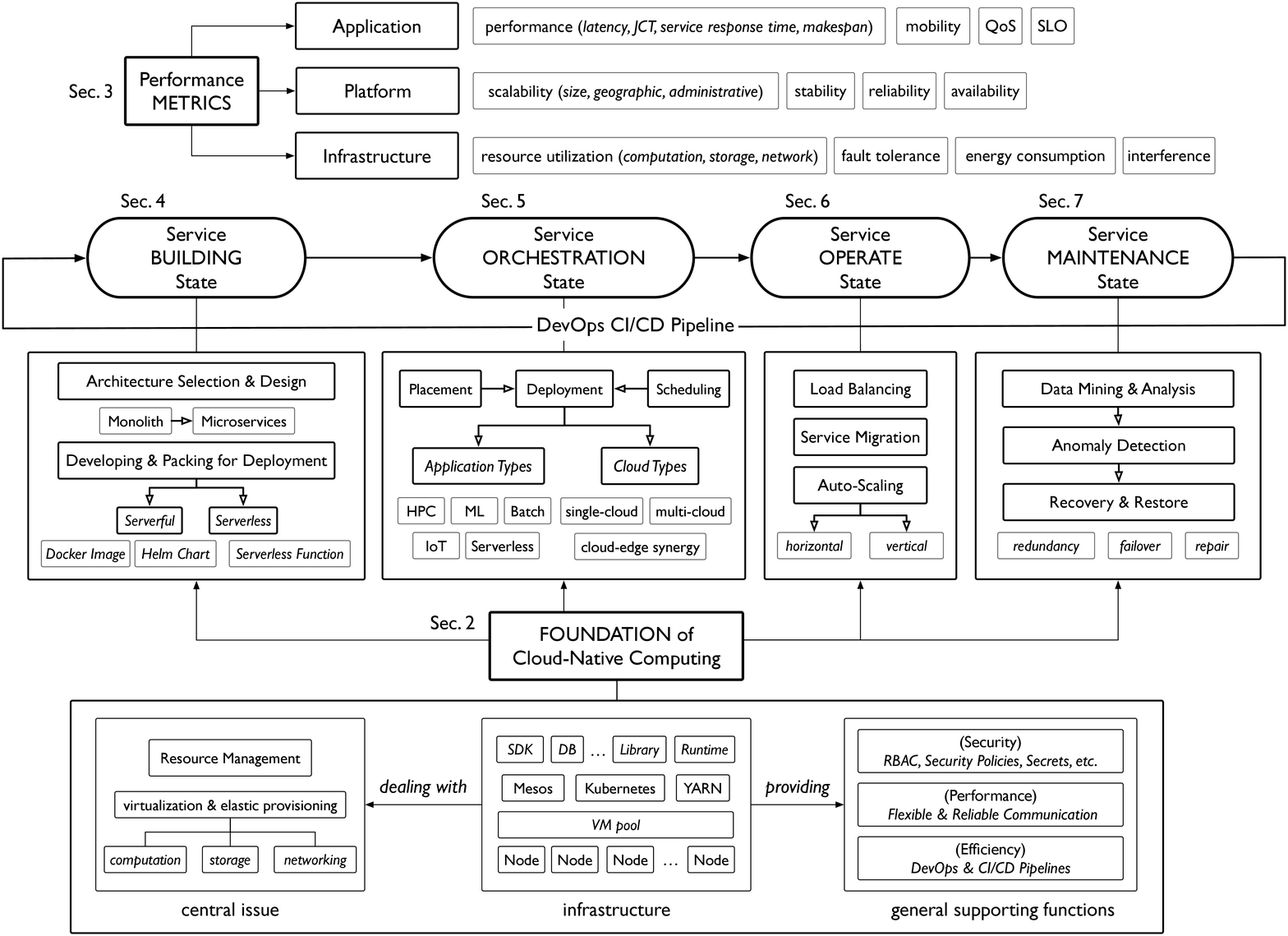}
  \caption{The research roadmap for cloud-native computing, designed from the perspective of services.}
  \label{f3}
\end{figure*}

As a conclusion, a cloud-native application can be viewed as a distributed, elastic and horizontal scalable system composed of inter-related microservices, which isolates state in a minimum of stateful components \cite{zhl-cloud-native-survey1}. Applications are built with cloud-native technologies by the following steps: $(i)$ Separating the monolith into self-deployed, function-explicit microservices and letting them communicate with each other through REST APIs (for synchronous communication) and lightweight messaging protocols (for asynchronous communication); $(ii)$ Using lightweight operating system virtualization technology, i.e., containerization, to pack each microservice into a container; $(iii)$ Orchestrating these containers into a organic whole for functionalities with automatic configuration and management throughout their life-cycle; $(iv)$ Using DevOps and CI/CDs to deliver the cloud-native applications with reliability and scalability. 

Considering that cloud-native is better known in industry, in this paper, we try to survey the past and present of cloud-native applications w.r.t. the key problems during their life-cycle from a research perspective. We attempt to merge the industrial popularity, including the widely used open-source software and platforms, with the trending researches, either theoretical or systematic, from the perspective of services computing. We divide the life-cycle of a cloud-native application, which is viewed as a \textit{service} in the field of service-oriented computing (SOC), into four states: \textit{building}, \textit{orchestration}, \textit{operate}, and \textit{maintenance}. As Fig. \ref{f3} shows, different key problems are emphasized in different states. In addition, we also collect the performance metrics which are frequently mentioned when building the cloud-native applications, and analyse them from three levels: infrastructure, platform, and software. Apart from the service life-cycle and the performance metrics, Fig. \ref{f3} also demonstrates the foundation of cloud-native computing. In our opinion, the fundamental issue to be dealt with for building cloud-native applications is resource management, i.e., resource virtualization and elastic provisioning, such that business agility can come true. Besides, as the foundation, it should provide general function modules for the building and running of cloud-native applications. We divide the functions into three cases: security, performance, and efficiency, and demonstrate the corresponding popular platforms and tools. The following sections will be presented surrounding the roadmap. 

\begin{table*}[htbp]   
  \vspace{-0.15cm}
  \begin{center}
  \caption{\label{tab-summary}Summary of important Acronyms.}   
  \begin{tabular}{c|c|c|c}    
      \toprule
      {\textsc{Acronym}} & {\textsc{Definition}} & {\textsc{Acronym}} & {\textsc{Definition}}\\[+0.1mm]
      \midrule
      \textbf{SOC} & Services-Oriented Computing & \textbf{SOA} & Service-Oriented Architecture \\[+0.7mm]
      \textbf{DevOps} & Combination of development and operations & \textbf{CI/CD} & continuous integration \& continuous delivery \\[+0.7mm]
      \textbf{CNCF} & Cloud Native Computing Foundation & \textbf{API} & application programming interfere \\[+0.7mm]
      \textbf{SDN} & software-defined network & \textbf{NFV} & network funciton virtualization \\[+0.7mm]
      \textbf{CNF} & cloud-native network function & \textbf{VXLAN} & Virtual eXtensible Local Area Network \\[+0.7mm]
      \textbf{GRE} & Generic Routing Encapsulation & \textbf{MPLS} & Multiprotocol Label Switching \\[+0.7mm]
      \textbf{LVM} & logical volume manager & \textbf{RAID} & redundant array of independent disks \\[+0.7mm]
      \textbf{CNI} & container network interface & \textbf{CSI} & container storage interface \\[+0.7mm]
      \textbf{SDK} & software development kit & \textbf{TLS} & transport layer security \\[+0.7mm]
      \textbf{NAT} & network address translation & \textbf{SaaS} & Software as a Service \\[+0.7mm]
      \textbf{PaaS} & Platform as a Service & \textbf{IaaS} & Infrastructure as a Service \\[+0.7mm]
      \textbf{JCT} & job completion time & \textbf{CRUD} & create, read, update and deletion\\[+0.7mm]
      \textbf{REST} & Representational State Transfer & \textbf{HPC} & high-performance computing \\[+0.7mm]
      \textbf{GA} & Genetic algorithm & \textbf{QoE} & Quality of Experience \\[+0.7mm]
      \textbf{QoS} & Quality of Service & \textbf{SLO} & Service-Level Objective \\[+0.7mm]
      \textbf{ANN} & Artificial Neural Network & \textbf{MARL} & Multi-Agent reinforcement learning \\[+0.7mm]
      \textbf{VM} & Virtual Machine & \textbf{ISP} & Internet Service Provider \\[+0.7mm]
      \bottomrule   
  \end{tabular}  
  \end{center}
  \vspace{-0.15cm}
\end{table*}

To the best of our knowledge, this is the first survey focusing on key issues during the life-cycle of cloud-native applications from the perspective of services. There are some studies on the origin, status quo, challenges and opportunities of cloud-native applications \cite{zhl-cloud-native-survey1,zhl-kratzke2018brief,zhl-gil2021cloud}. However, they mainly provide high-level opinions and ignore the review on the state-of-the-art research works. As a result, researchers may find it struggling to grasp and comprehend each issue in the development and management of cloud-native applications. It is worth mentioning that, there are surveys focusing on specific topics in cloud-native computing. For example, the author of \cite{zhl-duan2021intelligent} discusses the recent developments of architectural frameworks for intelligent and autonomous management for cloud-native networks. This paper comprehensively reviewed the technical trend toward cloud-native network design and network-cloud/edge convergence. Moreover, the scheduling of microservices and containers, which has a strong connection with the orchestration platform Kubernetes, is reviewed in \cite{zhl-senthuran2020review} and \cite{zhl-carrion2022kubernetes}. However, in the lack of systematic knowledge, challenges and proposed solutions will lack high portability and compatibility for various cloud-native applications. To this end, this survey is inspired to propose a \textit{pipelined} design and summarize the research domains from different views. It can help researchers and practitioners to further understand the nature of cloud-native applications. As shown in Fig. \ref{f3}, we analyze the key problems across the whole life-cycle of cloud-native applications, from the building state to the maintenance state. We also discuss the performance metrics and the fundamental necessities when developing cloud-native applications.

The rest of the survey is organized as follows: Sec. \ref{s2}-\ref{s7} introduce the foundation, performance metrics, and the four states of demonstrated in this roadmap. Sec. \ref{s8} discusses the issues, challenges, limitations, and opportunities of cloud-native. We conclude this paper in Sec. \ref{s9}. We summarize the definitions of the acronyms that will be frequently used in this paper in Table \ref{tab-summary} for ease of reference.


\section{Foundation of Cloud-Native Computing}\label{s2}
In this section, we demonstrate the fundamental necessities in cloud-native. From the perspective of clusters, we demonstrate the central issue of the foundation and general function modules which play critical roles when building cloud-native applications. 

\subsection{The Hierarchical Structure of Infrastructure}\label{s2.1}
A cluster is a set of computing nodes which are connected to each other through fast local area networks. On top of the physical nodes, OS-level virtualization techniques are utilized to create virtual machines (VMs) such that operating costs and downtime can be minimized. By maintaining a pool of VMs, fast provisioning of resources can be realized with cluster management softwares. Cluster management is always tightly coupled with resource management and task scheduling. Kubernetes, as we have mentioned before, is widely adopted across industries and has become a \textit{de facto} standard. Kubernetes has the ability of managing nodes (including both physical servers and VMs) with the module \textit{Node Controller}\footnote{https://kubernetes.io/docs/concepts/architecture/nodes/}, which is responsible for node registration, keeping the nodes up-to-date, and monitoring their health. In addition to Kubernetes, Mesos \cite{zhl-mesos}, Docker Swarm \cite{zhl-naik2016building}, and Hadoop YARN \cite{zhl-vavilapalli2013apache} are also widely used cluster managers. 
\begin{itemize}
  \item \textit{Kubernetes}. Kubernetes, abbreviated as K8s, is the most influential open-source platform in cloud-native since its release in 2014. K8s originates in Google's Borg \cite{zhl-verma2015large}, which has been used for managing containerized workloads in Google's inner clusters for more than a decade. K8s is a distributed software deployed across nodes, whose target is the automation of the deployment, management, and scaling of containerized applications by efficiently managing heterogeneous resources\footnote{The architecture of Kubernetes can be found in the official document: https://kubernetes.io/docs/concepts/overview/components/}. With the centre being K8s, an open-source ecosystem gradually forms to improve and facilitate the management of cloud-native applications. In the ever-growing K8s ecosystem, publishing, installing, removing, and upgrading of cloud-native applications can be managed with Helm\footnote{https://helm.sh/}. The traffic between internal microservices within an application can be managed with Istio\footnote{http://istio.io/}, which is the most influential implementation of Service Mesh\cite{zhl-li2019service}. Istio extends K8s to establish a programmable, application-aware network by taking Envoy\footnote{https://www.envoyproxy.io/} as the proxy. The monitoring of nodes and applications can be realized with Prometheus\footnote{https://prometheus.io} and Grafana\footnote{https://grafana.com} while logging can be managed with Kibana\footnote{https://www.elastic.co/kibana/}.
  \item \textit{Mesos}. Mesos is implemented in a two-level architecture. The first level is a \textit{master resource 
  manager} that dynamically controls which resources each \textit{framework scheduler} owns. Correspondingly, in the second level, each \textit{framework scheduler}, such as hadoop, MPI, and Mesos Marathon, are responsible for scheduling tasks at the application level \cite{zhl-mesos}.
  
  \item \textit{YARN}. YARN manages Hadoop clusters. It consists of clients, container, resource manager (RM), node manager (NM), and application master (AM). Here container is defined as a collection of physical resources such as RAM, CPU cores and disk on a single node. It is NM that takes care of each node and manages applications and workflows on that particular node. When a job is submitted by a client, the corresponding AM negotiates resources with RM and requests Container from the NM \cite{zhl-vavilapalli2013apache}.
  
  \item \textit{Docker Swarm}. Docker Swarm is the native inbuilt orchestration tool for Docker, which is called ``swarm mode''. Docker Swarm is composed of Docker Node, Docker Services, and Docker Tasks. There are two kinds of Docker Nodes, Manager and Worker, which are similar to the Master/Worker in K8s. The Manager, as the name suggests, is responsible for maintaining the cluster status, scheduling the services, and serving swarm mode HTTP API endpoints. By constrast, the Workers are nothing but the instances of Docker Engine for running Docker containers \cite{zhl-soppelsa2016native}.
\end{itemize}

The following contents are mainly focusing on K8s since in cloud-native, many rigor progress are achieved based on K8s. Apart from the above cluster managers, it is worth pointing out that there are orchestration frameworks that extend the native capabilities of K8s to the network edge. KubeEdge\footnote{https://kubeedge.io/en/} is a representative one. Edge computing has a three-level hierarchy: \textit{Cloud-Edge-Device} \cite{zhl-8746691}. To take the full advantage of this hierarchy, KueEdge divides its components into two parts: CloudCore and EdgeCore. CloudCore $(i)$ handles the communication between it and \textit{api-server} of a K8s cluster and $(ii)$ communicates with the edge nodes. EdgeCore is responsible for $(i)$ communicating with CloudCore and $(ii)$ manages the containers, services that are deployed on the edge devices\footnote{The architecture of KubeEdge can be found in the official document: https://kubeedge.io/en/docs/kubeedge/}.  In a recent project, KubeEdge is reported to stably support 100,000 concurrent edge nodes and manage more than one million \textit{Pods} \cite{zhl-kubeedge-test}.

On top of cluster managers, development tools including SDKs and middlewares are developed as the building blocks for cloud-native applications. The hierarchical structure of hardwares, OS-level softwares, cluster managers, and middlewares construct the fundamental necessities for building cloud-native applications. 

\subsection{Resource Provisioning and Management}\label{s2.2}
Virtualization refers to a collection of techniques for building and managing virtual resources on top of actual hardware, with key benefits including high redundancy, unified interfaces for users, and highly efficient resource utilization. It is the infrastructural foundation of today's cloud-native orchestration at all scales. When provisioning resources at a large scale, virtualization manages all low-level and possibly heterogeneous resources, such that better global efficiency can be reached in comparison to the traditional way of letting service users decide on their own, since service users often do not have access to the global resource utilization information. Typical computation virtualization technologies are summarized in Table \ref{cvt}.

There are multiple types of virtualization in the computing field, among which computation virtualization, storage virtualization, and network virtualization are most commonly used in cloud-native design and deployment. Due to the diversity of resources that can be virtualized, orchestrating all these heterogeneous types of devices is challenging. In this section, we will review virtualization technologies of different resources, their latest development, as well as the role they play when building cloud-native applications.

\subsubsection{Computation Virtualization}
Computation virtualization refers to creating an abstraction layer over computation, often in the form of virtual machines or containers. This is the core of all cloud-native deployments. The properties and efficiency of a specific virtualization technology deeply influence the entire deployment. 

\begin{table*}[htbp]   
  \vspace{-0.15cm}
  \begin{center}
  \caption{\label{cvt}Computation Virtualization Technologies.}   
  \begin{tabular}{c|c|c|c|c}    
      \toprule
      {\textsc{Name}} & {\textsc{Advantages}} & {\textsc{Disadvantages}} & {\textsc{Comments}} & {\textsc{Examples}}\\[+0.1mm]
      \midrule
      Type 1 hypervisor & \makecell{High efficiency,\\unmodified OS} & Low flexibility & \makecell{Also called "bare metal"} & \makecell{VMware ESXi,\\Microsoft Hyper-V}
      \\[+3mm]
      Type 2 hypervisor & \makecell{High compatibility,\\high flexibility,\\unmodified OS} & \makecell{Slightly lower efficiency\\than Type 1 hypervisor} & \makecell{Runs on OS} & \makecell{Oracle VirtualBox,\\VMware Workstation}
      \\[+3mm]
      Container & \makecell{Low overhead,\\flexible deployment} & \makecell{No vendored kernel} & \makecell{Foundation of cloud-native systems} & \makecell{Docker,\\FreeBSD jail}
      \\[+3mm]
      \bottomrule
  \end{tabular}  
  \end{center}
  \vspace{-0.15cm}
\end{table*}

From the perspective of where the hypervisor resides, virtualization technologies can be divided into Hypervisor Type 1 and Type 2. A hypervisor is a software layer that controls the creation and execution of VMs. Type 1 hypervisors, a.k.a. “bare-metal” hypervisors, directly run on hardware, while Type 2 hypervisors rely on an OS. Both hypervisors support unmodified guest OSs. Since Type 1 hypervisors directly communicate with hardware, they offer better performance and efficiency. Type 2 hypervisors, on the other hand, offer the best flexibility and compatibility, at the cost of a small portion of performance loss.

Another commonly used hypervisor in today's server hosting industry is KVM \cite{zc-KVM}. It is worth noting that KVM cannot be simply put into Type 1 or Type 2 hypervisor. While KVM runs in the Linux kernel and turns the kernel into a Type 1 hypervisor, the entire set of solution does operate on an existing operating system, making it Type 2 by definition. 

Instead of running the entire OS, containers choose another approach to virtualize computation resources. This new approach has been highly successful in today's cloud-native scenarios due to its high efficiency\cite{dyn-morabito2015hypervisors}. Containers share the OS kernel with the host OS but have a dedicated userland filesystem. Moreover, the filesystem only contains necessary binaries, libraries, and resource files, so the final image could be as low as a few hundred kilobytes. In many applications, having a completely isolated environment and a dedicated kernel is, in fact, a huge overkill. Currently, most cloud-native orchestration implementations, including Kubernetes, are based on Docker or other container engines\cite{Docker, dyn-kushwaha2017container}. This is mainly due to several unique advantages containers possess: simplicity, low overhead, fast deployment, and ease to design and build.

\subsubsection{Network Virtualization}
Network virtualization is another important level of virtualization in cloud-native scenarios. With network virtualization, traditional switches and routers are replaced with programmable devices, therefore allowing smarter operation. In this section, we will make a brief introduction to several key network virtualization technologies in the cloud-native context, including software-defined network (SDN), network function virtualization (NFV), Service Mesh, and overlay networks, to understand how they enable efficient and intelligent network management.

Software-defined Network (SDN) includes a set of techniques to decouple the data plane and control plane to enable programmatical a dynamical management of network forwarding devices like routers and switches. Traditionally, network operators leverage white-label devices from vendors to run their networks. This approach does not fit current quickly developing cloud-native environments due to the inflexibility and high price of vendor-made devices. A typical SDN system consists of several controllers and more forwarders. Traditional route or forward tables are replaced by unified flow tables, which are dynamically calculated by controllers and sent to forwarders. Forwarders then simply forward or drop traffic by looking up relevant table items from flow tables. The logically centralized control plane provides good visibility of the entire network, easing the management of network resources \cite{dyn-rawat2016software}. 

Network Function Virtualization (NFV) is another layer of virtualization in networking technologies. While SDN decouples the data plane and control plane, NFV focuses on decoupling software and hardware. With the quick development of computation virtualization, using virtualized software to replace network devices has become possible. Together with SDN, there have been several solutions, including firewall\cite{dyn-deng2015vnguard} and router\cite{dyn-harrabi2015implementing}. Furthermore, Cloud-native Network Function (CNF), a new cloud-native aware trend of Virtual Network Function has emerged. It is designed to run in containers instead of VMs, with the advantages of cloud-native fully leveraged. To sum up, By utilizing NFV and SDN, network operators like ISPs and cloud computing companies will benefit from reduced cost and improved flexibility to keep up with today's fast-evolving cloud-based trends.

\subsubsection{Storage Virtualization}
Storage virtualization is the technique of creating an abstraction layer over storage devices, to provide large, fast and redundant storage pools across multiple hard disks. As I/O operations take a large portion of the entire turnaround time, storage virtualization deeply affects the efficiency of the entire system. Furthermore, data redundancy and safety is an inherent requirement of cloud-native systems, which is often offered by storage. We identify three major layers of storage virtualization: Host-based Virtualization, Storage Device-based Virtualization, and Network-based Virtualization.

Host-based Virtualization is building the storage pool on the end host. An example of host-based virtualization is Logical Volume Manager (LVM). LVM is installed onto the OS, and creates a storage pool using storage devices connected to the host. Device-based Virtualization moves the virtualization layer from the OS to the device itself. A well-known example of this is Redundant Array of Independent Disks (RAID). There has been multiple combinations of RAID technologies (e.g. RAID-0, RAID-1, RAID-10, and RAID-60), for different requirements in regard to speed, redundancy and other specialized needs. Finally, Network-based Virtualization is often used in data centers. Network-based storage pool is built as a dedicated cluster of storage devices, and is connected to the end host using fast network links. As the storage cluster often live in the same data center as the hosts, optimal performance can still be achieved. All these three solutions do not need any modification on high level applications, as they have the same behavior as regular disk partitions. Therefore, good compatibility can be guaranteed.

To integrate low level storage systems into containers, Container Storage Interface (CSI)\footnote{https://kubernetes-csi.github.io/docs/} is proposed and introduced since Kubernetes v1.9. The CSI is a standard of exposing arbitrary low level storage system to containerized applications orchestrated by Cloud Orchestration systems like Kubernetes. To start using a new type of storage system, developers of the storage system are able to create a CSI plugin, without modifying the core of the Cloud Orchestration system in use. This is helpful in today's customized cloud-native deployments.

\subsection{General Supporting Function Modules}\label{s2.3}
In this section, we demonstrate the general function modules provided by the K8s ecosystem that play critical roles in the building and managing of cloud-native applications.

\subsubsection{Security}\label{s2.3.1}
K8s is designed with several security mechanisms to ensure the safety and confidentiality of data and resources within the cluster.
\begin{itemize}
  \item \textit{Role-Based Access Control (RBAC)}. RBAC\footnote{https://kubernetes.io/docs/reference/access-authn-authz/rbac/} is a security feature in K8s that enables system administrators to define specific access levels and permissions for each user or group of users within the cluster. With RBAC, it's possible to restrict certain privileges to only authorized entities.
  \item \textit{Pod Security Policies}. K8s provides Pod Security Policies that restrict the behavior of containers running inside the pods. These policies can prevent containers from executing privileged actions and running as root. By default, k8s also isolates pods from one another, which adds an additional layer of protection.
  \item \textit{Secrets Management}. K8s offers a facility, named as secret\footnote{https://kubernetes.io/docs/concepts/configuration/secret/}, for securely storing and managing sensitive data like passwords, certificates, keys, etc. The secret data is encrypted at rest and in transit, and the access to this data is restricted using RBAC.
\end{itemize}

\subsubsection{Performance}\label{s2.3.2}
Since the performance of containerization is mainly guaranteed by the underlying container engines, here we mainly discuss the performance of pod-to-pod (container-to-container, pod-to-service, etc.) communications. 

While SDN and NFV offer flexible network environments, they focus more on low-level communication, which is not easy to integrate with microservices. Ideally, microservices should focus on application logic, rather than low-level communications. Additionally, with hundreds even thousands of microservices cooperating with each other, it is harder to manage network communications as the number grows. K8s's inbuilt network support is able to provide basic network connectivity. Nevertheless, it is more common to use third-party network implementations that plug into K8s using the CNI (Container Network Interface) APIs. Typical implementations include Flannel\footnote{https://github.com/coreos/flannel}, Calico\footnote{https://github.com/projectcalico/cni-plugin}, Weave\footnote{https://www.weave.works/oss/net/}, etc. Flannel is the most popular implementation to configure a layer 3 network fabric for K8s. By using Flannel, each node will be installed a binary called \textit{flanneld}, which is responsible for allocating a subnet lease to each node out of a larger, preconfigured address space. The network built by Flannel uses VXLAN and many other cloud integrations for package forwarding \cite{zhl-vxlan}. Different implementations of CNI are compared in \cite{zhl-8539382}, \cite{zhl-9309003} and \cite{zhl-kumar2021networking}. K8s also has an inbuilt DNS such that \textit{Pods} and \textit{Services} can be discovered and visited though their domain names. 

\begin{figure}[htbp]
  \centering
  \includegraphics[width=2.2in]{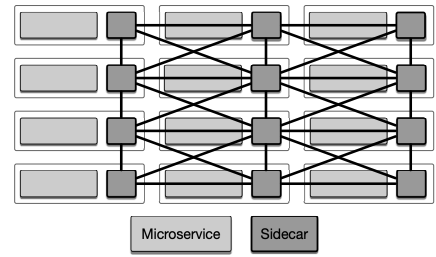}
  \caption{The sidecar proxies along with each microservice form a mesh-like network.}
  \label{mesh0}
\end{figure}

Note that the implementation of CNI is mandatory for building a working K8s cluster. However, to have a reliable, observable, and secure communication, the vanilla network is far from enough. Under the circumstances, service mesh is proposed, which is a software infrastructure working in \textit{the application layer} for controlling and monitoring internal, service-to-service traffic in microservice-based applications \cite{zhl-li2019service}. Fig. \ref{mesh0} illustrates why Service mesh is called a mesh. Service mesh provides dynamic discovery of services, intelligent load balancing across services, security features with encryption and authentication, and observability tracing by leveraging a so-called \textit{Sidecar} Design Pattern, where a sidecar proxy is dynamically injected into each \textit{Pod} for handling incoming requests. In service mesh, the control plane manages and configures proxies to route traffic, and collects and consolidates data plane telemetry. Correspondingly, the data plane is implemented as sidecar proxies. As we have mentioned before, Envoy is the most popular high-performance open-source implementation of sidecar proxy. 


\subsubsection{Efficiency}\label{s2.3.3}
DevOps, as a portmanteau of development and operation, is a collaborative and multidisciplinary effort within an organization to automate continuous delivery of new software versions, while guaranteeing their correctness and reliability \cite{zhl-3359981}. With DevOps, development and operations teams work together across the entire cloud-native applications' life-cycle. DevOps is strongly connected with CI/CD, which is a multi-stage pipeline of continuous integration (\textit{build $\to$ test $\to$ merge}), continuous delivery (\textit{automatically release the code to repository}), and continuous deployment (\textit{automatically deploy the application in production}). DevOps has already become the cornerstone of developing microservice-based cloud-native applications nowadays.

In the K8s ecosystem, KubeSphere DevOps is a powerful CI/CD platform that has attracted many attentions from industry \cite{zhl-ks}. KubeSphere DevOps provides CI/CD pipelines based on Jenkins\footnote{https://www.jenkins.io/}. It offers automation toolkits including Binary-to-Image (B2I) and Source-to-Image (S2I), and boosts continuous delivery across K8s clusters. A key character of KubeSphere DevOps is that it scales Jenkins Agents dynamically such that the CI/CD workflows can be accelerated flexibly.

\section{Performance Metrics}\label{s3}
By the top-down decomposition, the cloud native architecture is mainly composed of the application layer, the platform layer and the infrastructure layer. Different layers provides different type services from different perspective. First, at application layer, various software are deployed as different service that can be accessed by customers over network. The performance metrics related to the application contextual are latency, response time, completion time, makespan and so on. Then, platform layer provides the facilities and APIs to support the building and delivering of various services. The performance indicators related to the platform layer are scalability, stability, availability and so on. Finally, infrastructure layer provides the raw computing, storage, and network resources required by the service providers. The main performance metrics related to the infrastructure contextual are resource utilization, resource failure, energy consumption and so on. All in all, the service provision should be determined by single performance metric or jointly considering multi-criteria. We describe these performance metrics in detail as follows. 

\subsection{Performance Metrics in Application Level}\label{s3.1}
Various applications are encapsulated to be different containers and provides services to customers by deploying these containers to physical hosts in a cloud environment. Benefited from the convenience without installing and running the application programs. In application layer, the key problem is to guarantee the quality of service (QoS) for applications which can be measured by multiple performance metrics such as latency, response time, completion time, makespan. These performance metrics are very similar, and are often easily confused. But in fact, they have different meanings. In this paper, we distinguish these performance indicators and review these related work optimizing these indicators.

\begin{itemize}
    \item \textit{Latency}: it is usually referred to the time during between when something happens and when it is perceived. Some efforts presented in \cite{hbb-chen2022mobility, hbb-zheng2021multi} take latency as an important metric. For example, the authors in \cite{hbb-zheng2021multi} extend Kubernetes mechanisms to support multi-tenant at the cost of introducing the moderate latency and throughput overheads.
    \item \textit{Response Time}: it is referred to the time during from the request submission to the result return. Response time usually consists of the transmission time of data required by the request, the queuing time, the processing time and the result return time. Many works also consider to optimize response time. For example, the authors presented in \cite{hbb-wojciechowski2021netmarks} extend Kubernetes mechanism to schedule pod according to dynamic network metrics, the goal of which is to reduce the application response time. 
    \item \textit{Job Completion Time (JCT)}:  it is the time during from the start time of the entry task in job to the finish time of the last task. Different from response time, job completion time mainly concentrates on its processing time. Some efforts presented in \cite{hbb-fu2019progress, hbb-deng2021dependent, hbb-ogbuachi2019context, hbb-zhang2019multi} explicitly consider the job completion time. 
    \item \textit{Throughput}:  it is the ratio of processed requests to the total number of arrived requests at the system. A higher throughput indicates that much more requests are processed in unit time and much less response time are incurred by each request. Hence, optimizing throughput essentially is to optimize response time as well. Many works take the throughput maximization as an optimization objective \cite{hbb-yeung2021horus, hbb-han2021tailored}.
    \item \textit{Mobility}: the mobility of terminals also bring great challenges to service performance assurance. Specially, the service provision problem in the driver-less scenario has attracted extensive attention from academic and industry. Many works explore a series problems of service provision in terms of service placing, service scheduling, service migration, aiming at guaranteeing the service performance in the case of terminal moving. For instances, the authors presented in \cite{hbb-chen2022mobility} explicitly investigate the influence of the terminal mobility on service performance, and design some strategies to guarantee the service performance.
    \item {\textit{SLO}: Service Level Objective (SLO) is a key performance indicator that defines the level of service. It typically defined the minimum level of service that a provider must deliver to its customers and can be used to set expectations and establish accountability. A number of works presented in \cite{hbb-pusztai2021slo, hbb-pusztai2021novel} tries to guarantee SLO for an application.}
\end{itemize}

\subsection{Performance Metrics in Platform Level}\label{s3.2}
Platform as a service provides all facilities and APIs to build and deliver various services conveniently, which efficiently avoid tedious overhead incurred by downloading and installing the required software. The metrics to measure the platform service mainly include scalability, stability, reliability and availability. These metrics characterize the performances of platform service from different perspective. We describe these metrics and review these related studies optimizing these indicators in detail.

\begin{itemize}
    \item \textit{Scalability}: it is the ability of a system enabling to dynamically adjust the amount of resources allocated to containerized applications according to their potential workload fluctuations, which ensures the applications supported with enough resources to minimize SLA violations. Scaling can be performed vertically, horizontally, or both \cite{hbb-imdoukh2020machine}. At present, some efforts presented in \cite{hbb-imdoukh2020machine, hbb-pinciroli2020cedule+, hbb-jawaddi2022review, hbb-wang2019jily, hbb-nunes2021state} explicitly consider horizontal autoscaling, vertical autoscaling, and both. For example, in \cite{hbb-nguyen2020horizontal}, it can create much more container replicas to meet more application resource requirement. In \cite{hbb-liu2022coordinating}, it reallocates the amount of resource to the existing containerized application to best utilize the new hardware resource capacity. 
    \item \textit{Stability}: it is a system's ability to keep a quantity of required properties (e.g., queue length, waiting time, etc) within a bounded region when the system encounter some disturbances. To guarantee the stability of a system is a fundamental issue to ensure service performance. Therefore, some works investigate system stability problem and design various container deployment or placement approaches to guarantee service performance. For example, in \cite{hbb-rossi2019horizontal}, an adapted reinforcement learning algorithm is adopted to achieve horizontal and vertical elasticity of cloud application for increasing the flexibility to cope with varying workloads and guarantee performance stability. In \cite{hbb-rossi2020geo}, a two-step algorithm is designed to solve the container deployment problem in a geo-distributed computing environment. In the first step, a reinforcement learning approach is adopted to dynamically controls the number of replicas of individual containers on the basis of the application response time. In the second step, a network-aware heuristic algorithm is designed to place containers on geo-distributed computing resources. Its main goal is to satisfy Quality of Service requirements of latency-sensitive applications.
    \item \textit{Reliability}: it is referred to the system's ability to deliver services without service disruption, errors, or significant reductions in performance even when one or several of its software of hardware components fail. System reliability is also very important performance metrics. Many research efforts investigate the system reliability problem and design different software and hardware schemes to optimize this performance metric. To improve the reliability of the system and reduce makespan, a heuristic algorithm is proposed to balance load among virtual machines in \cite{zc-ebadifard2020dynamic}. To maintain reliability and elasticity for the system, a dynamic scheduling algorithm is proposed to balance the workload of virtual machines in a cloud environment elastically based on resource provisioning and de-provisioning methods. The above works mainly concentrate on solving software component failure to guarantee system reliability. Different from these above works, other works concentrate on solving hard component failure problem to guarantee system reliability. The author presented in \cite{de2020deep} predict the disk drives failure and overlap the time of regular data operation and data restoring to significantly improve service reliability and reduce data center downtime. 
    \item \textit{Availability}: it is the proportion of time a system is in a functioning condition. Along with scalability, stability, reliability, availability is also a prevailing issues for platform service. To cope with possible failure caused by the mobility of parked vehicles and improve the service availability, the author presented in \cite{hbb-nguyen2020collaborative} design the dual cost and utility-aware heuristic algorithm to solve the problem of multi-replica task scheduling in a collaborative computing paradigm consisting parked vehicles.
\end{itemize}

\subsection{Performance Metrics in Infrastructure Level}\label{s3.3}
Infrastructure as a Service provides the raw computing, network and storage resources and corresponding operating middleware software to customers on demand. One of the main benefits for infrastructure as a Service is free from the burden of infrastructure maintenance. In contrast to application as a service and platform as a service, infrastructure as a Service mainly provide to the resource service at lowest level. The performance metrics to measure the resource service mainly include resource utilization (computation, storage, network), failure rate, interference, or energy. We describe these performance metrics and review these related researches optimizing these indicators.

\begin{itemize}
     \item \textit{Resource Utilization}: It is an important performance metric used to describe the percentage of a system available resource, such computation resource, storage resource and network resource, that is occupied over an amount of available time (or capacity). In recent years, some efforts presented in \cite{hbb-bao2019deep, hbb-beltre2019kubesphere, hbb-bestari2020dynamic, hbb-carvalho2021qoe, hbb-toka2021ultra} explore resource planning and resource scheduling problems with the maximization of the CPU and RAM utilization. Specifically, the authors presented in \cite{hbb-beltre2019kubesphere} extend the Kubernetes mechanisms to fairly allocate multi-resource (such as CPU, memory, and disk) for containerized workloads of multi tenants. In \cite{hbb-warke2018storage}, a storage service orchestration platform is designed and implemented to support the stateful applications. In \cite{hbb-santoro2017foggy}, a workload orchestration framework is proposed to match infrastructure owner and tenants, aiming at optimizing the use of infrastructure while satisfying the application requirements. Not only that, but the network traffic is also taken as an optimization indicator \cite{hbb-caminero2021quality, hbb-nguyen2020elasticfog, hbb-santos2019towards}. For example, the authors in \cite{hbb-santos2019towards} and \cite{hbb-santos2019resource} design a network-aware scheduler to automatically manage and deploy containerized applications, aiming at for reducing the network latency. 

     \item \textit{Interference}: In cloud native environments, various types of workloads are encapsulated in the form of containers. However, the isolation of container is weaker than that of virtual machine. Multiple containerized workloads (such as computing intensive, storage intensive) co-located on the same server can interfere with each other, which seriously affect system performance. The interference issue incurred by co-locating different type workloads become a pressing issue. Therefore, many works presented in \cite{hbb-fu2019progress, hbb-yeung2021horus, hbb-kaur2019keids, hbb-zhong2020heterogeneous, hbb-yim2021qos} explicitly consider the inference between co-locate containerized jobs. For instance, the authors propose in \cite{hbb-fu2019progress} a container placement scheme that balances the resource contention on the worker nodes.
    
     \item \textit{Energy Consumption}: It is the amount of energy used. The significant amount of energy consumed by data centers can incur high cost and environmental pollution. Moreover, the energy consumption problem is also very important for resource-constrained terminals, due to their limited battery capacity. In recent years, there exist research efforts on designing various energy-efficient schedulers \cite{hbb-kaur2019keids, hbb-chhikara2020efficient, hbb-xu2019brownout, hbb-gunasekaran2020fifer, hbb-thinakaran2019kube}. For example, the authors presented in \cite{hbb-chhikara2020efficient} propose an energy-efficient container migration scheme to migrate containers for reducing the energy consumption. The authors presented in \cite{hbb-kaur2019keids} design a scheduler to minimize energy consumption and interference. 

     \item \textit{Cost}: Currently, the big infrastructure service providers such as AWS, Azure and Alibaba mainly adopt the pay-as-you-go payment model. Therefore, the financial costs for renting infrastructure resource is also very import performance metric. In recent years, some research efforts presented in \cite{hbb-chung2018stratus} \cite{hbb-zhong2020cost} \cite{hbb-xu2017cost} \cite{hbb-xu2020self} take financial costs as an optimization goal, and find an optimal orchestration solution by selecting diverse cloud services according to their pricing models and computing capability. Their main goals are to minimize the overall financial costs while satisfying the QoS requirements.

     \item \textit{Fault Tolerance}: It is the ability of a system to behave in a well-defined manner once faults occur. Failures could occur due to dynamic changes in the execution environment. The failures in the IaaS layer or the physical hardware has heavily negative effect to the system. Hence, it is import to design various fault tolerance mechanism to cope with this problem and to minimize the risk of failure. For instances, in \cite{hbb-kim2019overlit}, the authors develop a new container storage driver to solve the global failures and bundled performance problem.
\end{itemize}


\section{Service Building}\label{s4}

In service building state, the key steps are $(i)$ architecture selection and $(ii)$ code development \& packing. 

\subsection{From Monolith to Microservices}\label{s4.1}
Before the rise of the microservices architecture, many traditional applications adopt monolithic architectures. In this case, the application is deployed in the shape of a single-tiered monolith, which combines different components into a single program. Typical components are listed as follows:
\begin{itemize}
    \item \textit{Business Logic}. The application's core business logic. For example, in e-commerce websites, the logic of inventory and shipping management.
    \item \textit{Database}. The data access objects responsible for the CRUD of data.
    \item \textit{Interaction and Presentation}. The component responsible for handling HTTP requests and responding with either HTML or JSON/XML (for web services APIs) objects.
    \item \textit{Integration}. The component responsible for the integration with other internal or external services though message protocols or REST APIs.
\end{itemize}
With monolithic architectures, all components are tightly coupled and run as a single service. As a result, any component of the application experiences a spike in demand, the entire architecture has to be scaled. Besides, adding or improving a monolithic application’s features becomes complex as the code base grows. This greatly increase the risk for availability since many dependent and tightly coupled components increase the impact of a single failure. 

To solve the above problems, the microservices architecture is proposed and it becomes the dominant architectural style choice for service-oriented software \cite{zhl-alshuqayran2016systematic}. With a microservices architecture, an application is built as independent components that run separately as a single service. These services communicate via a well-defined interface using lightweight APIs. Services are built for business capabilities and each service performs a single function module. Since they are independently run, each service can be updated, deployed, and scaled to meet demand for specific functions of an application \cite{zhl-gos2020comparison,zhl-de2019monolithic,zhl-ponce2019migrating}. The microservices architecture brings in many benefits, such as agility, flexible scaling, easy deployment, resilience, etc. When developing the cloud-native applications, the primary task is to select the appropriate architecture (monolith or microservice) based on specific business logic.

\subsection{Packing Microservices into Containers}\label{s4.2}
Microservices are usually packaged as container images using container technologies such as Docker and then published to an image registry. The most popular choice for deploying these container images on a container orchestration platform like Kubernetes is Helm. Helm charts contain references to the publicly accessible container registry in order to pull the necessary container images. Nevertheless, certain companies and organizations uphold their own private cloud infrastructure. In such cases, accessing public container image registries or the internet from within the private cloud is restricted. To deploy an application in such a limited environment, it becomes necessary to bundle all the required artifacts, including container images, Helm charts, documentation, etc., into an archive.

It worth mentioning that, if the cloud-native application is published through serverless functions, the developer only need to upload the code to the serverless platform, and the containerization and orchestration is automatically executed by the underlying middlewares and tools. Serverless computing is a method of providing backend services on an as-used basis \cite{zhl-jonas2019cloud}. A serverless provider allows users to write and deploy code without the hassle of worrying about the underlying infrastructure. A company that gets backend services from a serverless vendor is charged based on their computation and do not have to reserve and pay for a fixed amount of bandwidth or number of servers, as the service is auto-scaling. Note that despite the name serverless, physical servers are still used but developers do not need to be aware of them. Serverless computing allows developers to purchase backend services on a flexible ``pay-as-you-go'' basis, meaning that developers only have to pay for the services they use. Detailed reviews of recent works on serverless computing will be given in Sec. \ref{s5.1.3}.


\section{Service Orchestration}\label{s5}

Service orchestration is the automated configuration, management, and coordination of multiple microservices to deliver the end-to-end services. Since microservices are encapsulated in form of container, service orchestration is essentially container orchestration. As a popular open-source container orchestration tool, Kubernetes is able to automatically deploy a large number of containers, and coordinate them to work together in congruence, thereby greatly reducing operational burdens. The key technology to support service orchestration lies in effective service placement and dynamic service scheduling \cite{hbb-han2021tailored, hbb-carrion2022kubernetes}. In the cloud native context, these two key technologies are investigated widely by a plenty of studies. The service orchestration solution is mainly affected by the characteristics of the applications and the computational architectures. Hence, these  orchestration solutions can be classified based on the type of applications and the computational architectures. The subcategories match the following questions. Representative works are listed in Table \ref{tab-orchestrate}.  
\begin{itemize}
    \item What type of applications are orchestrated in cloud native system?
    \item What computational architectures are used in the service orchestration?
\end{itemize}

\begin{center}
\begin{table*}[]
  \caption{\label{tab-orchestrate}Representative works in service orchestration.}   
  \begin{tabular}{c|c|ccc|ccc|c|c}
    \toprule
    \multirow{2}{*}{\textsc{Work}} & \multirow{2}{*}{\textsc{App. Type}} & \multicolumn{3}{c|}{\textsc{Resource}}                    & \multicolumn{3}{c|}{\textsc{Cloud Type}}               & \multicolumn{1}{c|}{\multirow{2}{*}{\textsc{Metric}}} & \multirow{2}{*}{\textsc{Implementation}} \\
                          &                            & CPU            & Storage        & B.W.           & Single-Cloud   & Multi-Cloud & Cloud-Edge     & \multicolumn{1}{c|}{}                         &                                 \\ \midrule
                          \cite{hbb-wang2022preemptive} & \multirow{9}{*}{ML}        & \checkmark & \checkmark & \checkmark &                &             & \checkmark & JCT                                           & simulation                            \\
                          \cite{hbb-xu2020acceleration}                    &                            & \checkmark & \checkmark & \checkmark &                &             & \checkmark & JCT                                           & simulation                            \\
                          \cite{hbb-huang2019gpipe}                    &                            & \checkmark & \checkmark & \checkmark & \checkmark &             &                & training time                                 & simulation                            \\
                          \cite{hbb-wang2020job}                   &                            & \checkmark & \checkmark & \checkmark & \checkmark &             &                & JCT \& makespan                               & system                            \\
                          \cite{hbb-albahar2022schedtune}                   &                            &                & \checkmark &                & \checkmark &             &                & makespan                                      & system                            \\
                          \cite{hbb-wu2020irina}                   &                            & \checkmark &                &                & \checkmark &             &                & JCT                                           & simulation                            \\
                          \cite{hbb-shen2019nexus}                   &                            & \checkmark &                &                & \checkmark &             &                & latency \& res. utilization                   & simulation                            \\
                          \cite{hbb-yeung2020horus}                    &                            & \checkmark &                &                & \checkmark &             &                & inference time                                & simulation                            \\
                          \cite{hbb-zhou2021container}                   &                            & \checkmark & \checkmark &                & \checkmark &             &                & response time                                 & system                            \\ \midrule
                          \cite{hbb-misale2021s}                   & \multirow{6}{*}{HPC}       & \checkmark & \checkmark &                & \checkmark &             &                & response time                                 & system                            \\
                          \cite{hbb-misale2021towards}                   &                            & \checkmark & \checkmark & \checkmark & \checkmark &             &                & response time                                 & simulation                            \\
                          \cite{hbb-beltre2019enabling}                   &                            & \checkmark & \checkmark &                & \checkmark &             &                & througput                                     & simulation                            \\
    \cite{hbb-lopez2020seamlessly,hbb-choochotkaew2022autodeck}               &                            & \checkmark & \checkmark & \checkmark & \checkmark &             &                & response time                                 & system                            \\
                          \cite{hbb-deng2021dependent}                    &                            & \checkmark & \checkmark &                &                &             & \checkmark & makespane                                     & simulation                            \\
                          \cite{hbb-fan2020knative}                   &                            & \checkmark &                &                & \checkmark &             &                & prediction accuracy                           & simulation                            \\ \midrule
                          \cite{hbb-ling2019pigeon}                   & \multirow{6}{*}{serverless} & \checkmark & \checkmark & \checkmark & \checkmark &             &                & response time               & simulation                            \\
                          \cite{hbb-kaffes2019centralized}                   &                             & \checkmark & \checkmark &                & \checkmark &             &                & scalability                 & system                            \\
                          \cite{hbb-venkataraman2014power}                   &                             & \checkmark & \checkmark & \checkmark & \checkmark &             &                & JCT                         & simulation                            \\
                          \cite{hbb-venkataraman2017drizzle}                   &                             & \checkmark & \checkmark &                & \checkmark &             &                & latency \& throughput       & simulation                            \\
                          \cite{hbb-wang2019distributed}                   &                             &                & \checkmark &                & \checkmark &             &                & JCT                         & simulation                            \\
                          \cite{hbb-gu2022fluid}                   &                             & \checkmark & \checkmark &                & \checkmark &             &                & JCT \& res. utilization     & system                            \\ \midrule
                          \cite{hbb-zhang2021zeus}                   & \multirow{5}{*}{batch job}  & \checkmark & \checkmark &                & \checkmark &             &                & SLO \& res. utilization     & simulation                            \\
                          \cite{hbb-chen2020woa}                   &                             & \checkmark & \checkmark & \checkmark & \checkmark &             &                & res. utilization            & simulation                            \\
                          \cite{hbb-zhao2019distributed}                   &                             & \checkmark & \checkmark & \checkmark &                & \checkmark &  & JCT                         & simulation                            \\
                          \cite{hbb-ye2021shws}                   &                             & \checkmark & \checkmark & \checkmark & \checkmark &             &                & res. utilization \& cost    & simulation                            \\
                          \cite{hbb-hu2020concurrent}                   &                             & \checkmark & \checkmark & \checkmark & \checkmark &             &                & JCT \& res. utilization     & simulation                            \\ \bottomrule
  \end{tabular}
\end{table*}
\end{center}

\subsection{Application Types}\label{s5.1}
\begin{figure}[htbp]
  \centering
  \includegraphics[width=3.5in]{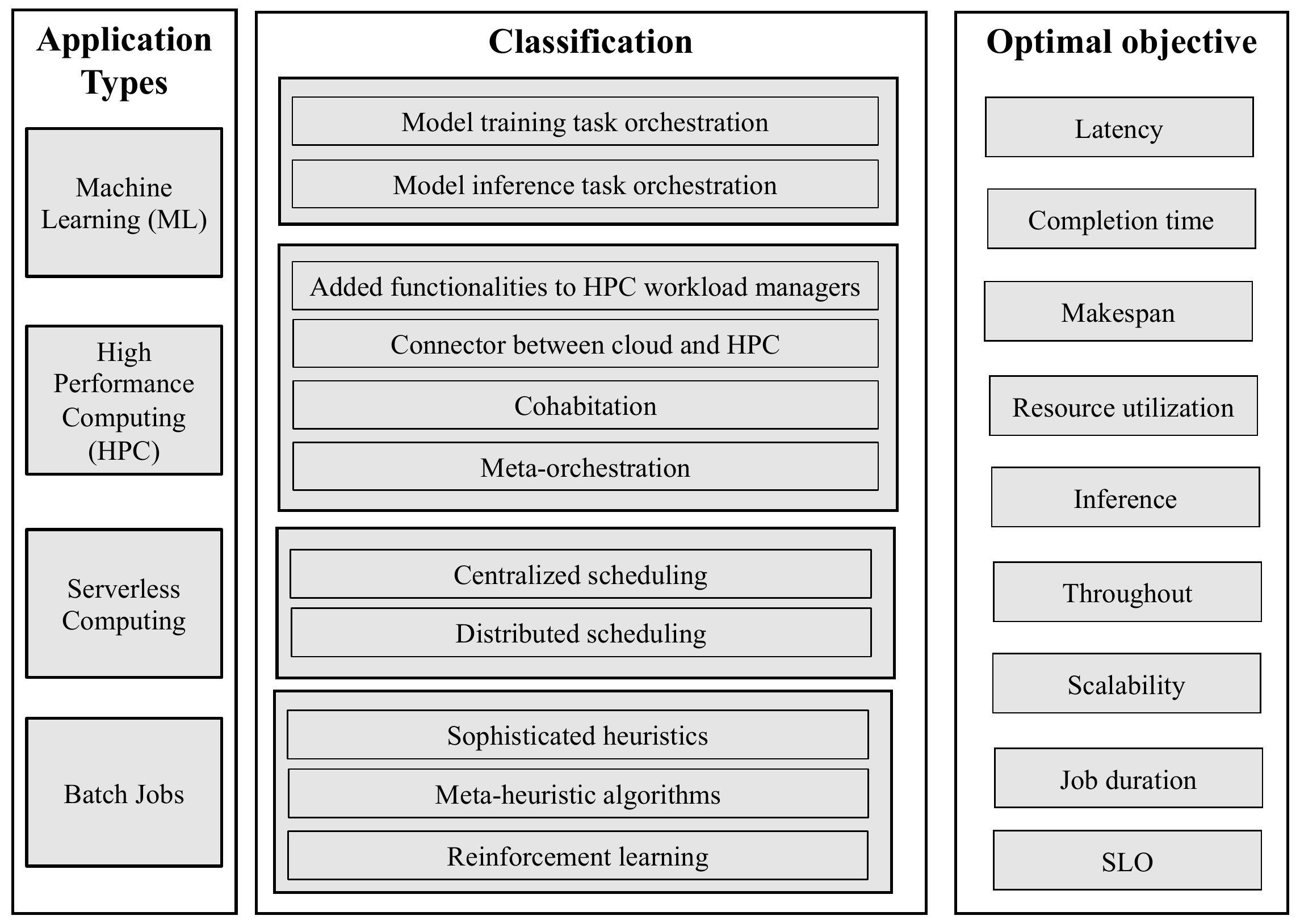}
  \caption{Service orchestration under different application types.}
  \label{serviceOrchestration}
\end{figure}

Different types application have significantly distinct characteristics, such as their Quality of Service requirements, the type, the structure and so on. These characteristics of the applications have a significant impact on the service orchestration. A wide range of applications from High-Performance Computing (HPC), machine learning, batch, web service or serverless, are handled in cloud native system. In the following, we review these research studies of service orchestration for different types of applications. Fig. \ref{serviceOrchestration} outlines the structure of this section.

\subsubsection{Machine Learning}\label{s5.1.1}
As a subfield of machine learning, deep learning has become a very popular research topic due to its advancement in various applications. A standard deep learning development pipeline consist of model training and model inference two stages. These two stage tasks are characterized by unique and complicated features. Specifically, the model training task is long-lived offline task, while the model inference task is short-lived online task. Moreover, these two stage tasks focus different performance metrics, in which the model training task focus on achieving high performance and the model inference task pay more attention on the response latency and inference accuracy. Their unique characteristics and different performance requirements impose some specific challenges to orchestrate model training tasks and model inference tasks. We comprehensively review and summarize these studies related to model training task orchestration and model inference task, respectively. 

\textbf{Model training task orchestration.} 
Model training is the process of learning a model over a large dataset using a machine learning algorithm. Due to increasingly complicated model and larger datasets, model training is an extremely time consuming and resource consuming task. Thus, it is urgent to parallelize model training task on multiple distributed workers. There are mainly four types of parallelism model training: data parallelism  \cite{hbb-yu2022gadget , hbb-mao2021speculative, hbb-zheng2019flowcon, hbb-zheng2019cynthia, hbb-wang2022preemptive}, model parallelism\cite{hbb-xu2020acceleration}, pipeline parallelism \cite{hbb-huang2019gpipe, hbb-narayanan2019pipedream, hbb-li2021respipe, hbb-luo2022efficient} and mixed parallelism \cite{hbb-yi2020fast, hbb-wang2020job}. Data parallelism refers to place multiple replicas of a model on multiple workers, and divide the datasets into many subsets to feed to these multiple workers. These multipe workers simultaneously perform the model training tasks, and synchronize their training results in the form of parameter servers or All-Reduce, etc. By data parallelism, the speed for model training can be greatly accelerated and the performance for model training can be enhanced. For example, in \cite{hbb-wang2022preemptive}, a novel online preemptive scheduling framework is designed to dispatch machine learning jobs to workers and parameter server for reducing the average job completion time. Analogously, in \cite{hbb-albahar2022schedtune}, the authors present a heterogeneity-aware scheduler which can efficiently collocate deep learning jobs on GPUs by exploiting the predicting information of GPU memory demand and job completion times. Its main goal is to improve the GPU resource utilization and reducing the makespan. However, with the increase of model complexity, a model with the large number of parameters cannot be launched on a single worker. Thus, model parallelism is proposed. Model parallelism refer to divide a model into multiple disjoint partitions and place these partitions on multiple workers. Since each worker only have one part of the model, only one worker is performing the model training task at any one time. One of the problems for the model parallelism is the long training latency incurred by the communication among multiple workers. To address this problem, a novel parallelism model training, called pipeline parallelism, is proposed. Pipeline parallelism divides a model into multiple stages and place multiple stages on multiple workers; Beside, pipeline parallelism further divide the datasets into multiple micro-batches. Multiple workers can process multiple micro-batches simultaneously. Thus, pipeline parallelism training greatly reduce the training latency. For example, to achieve an efficient and task-independent model parallelism, the authors in \cite{hbb-huang2019gpipe} introduce a pipeline parallelism library to partition a deep neural network across multiple workers and split the datasets into mini-batches. Finally, mixed parallelism with data and model parallelism is proposed to reduce the training latency and resource consumption. For example, in \cite{hbb-wang2020job}, the authors design a job scheduling system which enable machine learning jobs to be implemented with data parallelism and model parallelism in clusters. The proposed system greatly reduces the job completion time and improves accuracy.

\textbf{Model inference task orchestration.}
Model inference is the service ability to make predictions on new data based a trained model. Model inference tasks are usually deployed as online short-lived services (e.g., automatic driving, face recognition). How to deploy and orchestrate model inference tasks has attracted much attention in industry and academia. For industry, many mainstream deep learning frameworks, such as TensorFlow Serving \cite{hbb-olston2017tensorflow} and MXNet Model Server \cite{hbb-2022multi}, have implemented the orchestration function for model inference task. For academia, there are also a lot of studies about model inference task orchestration. These studies mainly can be classified two types: the individual model inference task orchestration and multiple model inference task orchestration. For the single model inference task orchestration, the authors in \cite{hbb-liang2021pruning, hbb-gholami2021survey} design some optimization techniques to efficiently orchestrate model inference task. However, the execution of a single model inference task not only fail to meet the requirement of application scenarios, but also causes the waste of the resources. Thus, many researchers further study multiple model inference task orchestration. They design different heuristic-based, modeling-based or prediction-based mechanisms to orchestrate multiple model inference tasks\cite{hbb-choi2020prema, hbb-mendoza2021interference, hbb-wu2020irina}. Specifically, the authors in \cite{hbb-shen2019nexus} adopt a heuristic approach to select the requests to be co-located on the same GPU. First, they assign the optimal batch sizes, given the latency requirement of existing inference task requests. Finally, they establish the node runtime cycles, aiming at maximizing the resource utilization while satisfying the latency requirement. Analogously, the authors in \cite{hbb-wu2020irina} colocate these model inference tasks where the total of their peak GPU requirement does not exceed the capacity, and heuristically schedule the newly arrived model inference task to the worker with smallest completion time, aiming at reducing the total delay. However, all of the above studies mainly exploit the heuristic method to orchestrate the certain deep learning model scenarios at a limited scale. The performance of these heuristic methods could dramatically degrade when the deep learning models vary. Thus, these heuristic methods can not be applied to cope with dynamic colocation mechanisms for managing the inference workloads. With complexity and dynamics of model inference task, many researchers turn to learning-based methods such as multi-armed bandit, reinforcement learning, etc. For instance, in \cite{hbb-yeung2021horus}, the authors investigate the job completion time slowdown problem caused by the interference between co-located deep learning jobs. To address this problem, an interference-aware resource manager is designed to effectively co-locate heterogenous deep learning jobs for improving resource utilization and job throughput.

\subsubsection{High Performance Computing}\label{s5.1.2}
 High Performance Computing (HPC) jobs are usually large workloads such as large-scale financial, scientific computing and engineering simulation. To execute these HPC jobs, an amount of computing power, memory and network speeds tend to be required. HPC jobs are often submitted to an HPC cluster and wait to be scheduled by a HPC job scheduler. However, the existing HPC job schedulers lack micro-service support and container management capacities. Therefore, it is a challenge about how to efficiently support HPC workloads on Kubernetes. In recent years, there exist research efforts on efficiently orchestrating HPC jobs on cloud clusters \cite{hbb-zhou2022containerisation, hbb-reuther2018scalable, hbb-netto2018hpc, hbb-fan2021job}. These state-of-the-art studies on HPC job orchestration can be divided to four categories: added functionalities to HPC workload managers \cite{hbb-wofford2020layered, hbb-julian2016containers, hbb-higgins2015orchestrating}, connector between cloud and HPC \cite{hbb-zhou2020container, hbb-zhou2021container, hbb-zhou2021containerization, hbb-zhou2022cybele}, cohabitation \cite{hbb-beltre2019enabling, hbb-liu2018dynamically, hbb-piras2019container}, meta-orchestration \cite{hbb-carnero2018running, hbb-di2020approach, hbb-colonnelli2020streamflow}. For added functionalities to HPC workload managers, it mainly extends HPC workload manager to support container orchestrator for HPC application. For example, in \cite{hbb-misale2021s, hbb-zhou2021container}, the authors investigate the problem of running HPC workloads efficiently on Kubernetes clusters, and implement a plug-in to efficiently schedule the HPC workloads. The benefit of this HPC job orchestration approach is less intrusive. However, its disadvantage is that the added functionalities are limited. To address the shortcoming for added functionalities to HPC workload managers, the connector between cloud and HPC is proposed. The connector enable to bridge the gap between HPC and cloud systems and achieve HPC job orchestration on cloud platform. For example, in \cite{hbb-misale2021towards}, a scheduler plug-in for Kubernetes is implemented to efficiently schedule the HPC applications on cloud platforms managed by Kubernetes. In \cite{hbb-di2017nextflow}, a workflow management system is implemented to schedule the containerized HPC applications such as nextflow, which greatly improve the numerical instability incurred by variations across computational platforms. The benefit of the connector is non-intrusive and enable to exploit orchestration strategies of orchestration platforms. However, its disadvantage is that the network latency between cloud and HPC is high. Therefore, a HPC job orchestration approach, called cohabitaion, is proposed. The cohabitaion is to coexist HPC workloads manager and cloud orchestrators on an HPC cluster. For instances, in \cite{hbb-beltre2019enabling}, the authors investigate the service orchestration problem towards HPC workloads. To address this problem, the authors modify the configuration and setup of Kubernetes to support HPC workloads, and evaluate the performance of HPC workloads. The cohabitaion has an advantage of fully exploiting the functionalities of orchestration platforms. However, the cohabitaion is extremely intrusive. Therefore, a HPC job orchestration approach, called meta-orchestration, is designed. The meta-orchestration approach is to implement an additional orchestrator on top of the cloud orchestrator and HPC workload manager. For example, in \cite{hbb-kubebatch}, a framework called Kube-batch is designed to enable HPC workloads execution on Kubernetes. In \cite{hbb-lopez2020seamlessly}, an open source tool is designed to manage the full life cycle of HPC workloads in cloud architectures. In \cite{hbb-choochotkaew2022autodeck}, a framework which is compatibility with Prometheus is proposed to automatically deploy the benchmarking workload for containerized HPC applications and analyze their performances. The advantage of the meta-orchestration approach is less intrusive. However, its disadvantage is to increase the complexity of the architecture and the efforts of maintenance.

\subsubsection{Serverless Computing}\label{s5.1.3}
Severless computing is a new execution model of cloud computing which is a integration of both function as a service (FaaS) and backend as a service (BaaS). Serverless computing is characterized by the automatic management and lightweight features. These characteristics of serverless computing enable developers to focus on the business logic, with no need worry about infrastructure provisioning and management. Benefiting from its advantages, serverless computing recently attracts a lot of attentions in both industry and academia. However, the inextricable dependencies between massive functions pose a great challenge to serverless orchestration. In recent years, there are a plenty of research efforts on serverless orchestration. In the industry community, several open-source platforms and serverless computing frameworks, such as Kubeless \cite{hbb-project2022kubeless}, OpenFaas \cite{hbb-govind2021benchmarking}, OpenWhisk \cite{hbb-djemame2020open} or Fission \cite{hbb-mohanty2018evaluation}, are designed to support the serverless computing orchestration. These open source frameworks with different architectures enable to dynamically manage, scale, and provide different types of resources for serverless applicaitons. In the academic community, some research efforts presented in  \cite{hbb-fan2020knative, hbb-ling2019pigeon} design diverse scheduling scheme for serverless applications. These strategies can roughly be divided two categories: centralized scheduling \cite{hbb-kaffes2019centralized, hbb-venkataraman2014power, hbb-venkataraman2017drizzle} and distributed scheduling \cite{hbb-wang2019distributed, hbb-mashayekhi2017execution}. For the centralized scheduler, the authors in \cite{hbb-fan2020knative} present a double exponential smoothing approach to calculate the optimal number of pods for serverless applications. Analogously, in \cite{hbb-ling2019pigeon}, a serverless computing frameworks, called Pigeon, is presented to schedule the FaaS funcation to pre-warmed containers. Moreover, the framework introduces a static pre-warmed container pool to cope with the burst function arrival. Both of novelty mechanism can greatly reducing the response time for serverless application and improve the system performance. Moreover, in \cite{hbb-deng2021dependent}, the authors investigate the influence of the composite property of services on scheduling scheme at serverless edge. To address this problem, a dependent function embedding algorithm is designed to get the optimal edge server for each function, aiming to minimize its completion time. All above these approaches are centralized. The centralized schedulers are vulnerable to a single point of failure and high communication overhead. To address these problem of the centralized scheduler, some distributed scheduling strategies are proposed. For example, the authors in \cite{hbb-wang2019distributed} design a scheduler based on deep reinforcement learning to dynamically make decision on the number of functions and their resources, aiming at making a trade-off between cost and performance.

\subsubsection{Batch Jobs}\label{s5.1.4}
More and more diverse tasks are running on cloud data centers, of which batch jobs account for a large proportion. There exist many works dealing with batch job scheduling. These works are mainly carried from the system implementation and algorithm optimization two aspects. For the works on system implementation, the authors in \cite{hbb-gu2022fluid}, design a cloud-native platform called Fluid which can co-orchestrate the data cache and deep learning jobs to improve the overall performance of multiple deep learning jobs. Analogously, in \cite{hbb-zhang2021zeus}, a scheduling system based on the real server utilization and a sliding window-based algorithm are designed to schedule and reschedule batch jobs, and thereby effectively improve the resource utilization in Kubernetes. For the works on optimization algorithm, there are mainly three kinds of methods to solve it: sophisticated heuristics, meta-heuristic algorithms \cite{hbb-chen2020woa, hbb-zhao2019distributed, hbb-ye2021shws, hbb-hu2020concurrent} and reinforcement learning \cite{hbb-huang2020rlsk, hbb-gu2021characterizing}. The sophisticated heuristics, such fair scheduling \cite{hbb-hadoop}, first-fit \cite{hbb-song2013adaptive}, simple packing strategies \cite{hbb-grandl2014multi} are usually easy to understand and implement. However, it needs manual adjustment to gradually improve the algorithm. Therefore,the meta-heuristic algorithms, such genetic algorithm or ant colony algorithm, are proposed to orchestrate batch jobs. For example, in \cite{hbb-chen2020woa}, the authors adopt the meta-heuristics optimization algorithm to schedule batch jobs for achieve higher resource utilization. In \cite{hbb-zhao2019distributed}, a redundant placement problem for microservice-based applications is formulated to be a stochastic optimization problem. To address this problem, a GA-based server selection algorithm is designed to efficiently decide about how many instances as well as on which edge sites to place them for each microservice. Its main goal is to reduce service execution latency and improve the service availability. Analogously, in \cite{hbb-ye2021shws}, a stochastic hybrid workflow scheduling algorithm is design to jointly schedule offline batch workflows and online stream workflows in cloud container services. Its main goal is to minimize the cost and improving resource utilization in cloud container services. Moreover, in \cite{hbb-hu2020concurrent}, the authors formulate the concurrent container scheduling problem to be a minimum cost flow problem. To address this problem, an efficient solution is designed to lower the average container completion time and improve resource utilization. However, the batch job orchestrations based on meta-heuristic algorithms can not efficiently cope with the dynamics of the batch jobs and the variety of the execution environment. To address this problem, reinforcement learning methods are adopted to handle dynamic orchestration problems of batch jobs. For instances, in \cite{hbb-huang2020rlsk}, the authors adopt a deep reinforcement learning algorithm to schedule independent batch jobs among multiple clusters adaptively. Analogously, the authors in \cite{hbb-gu2021characterizing} propose a graph learning approach to discover the insightful properties and patterns of batch jobs. Based on these characteristics, the batch jobs can be better scheduled in production cloud computing environment.

\subsection{Cloud Types}\label{s5.2}
\begin{figure}[htbp]
  \centering
  \includegraphics[width=3.5in]{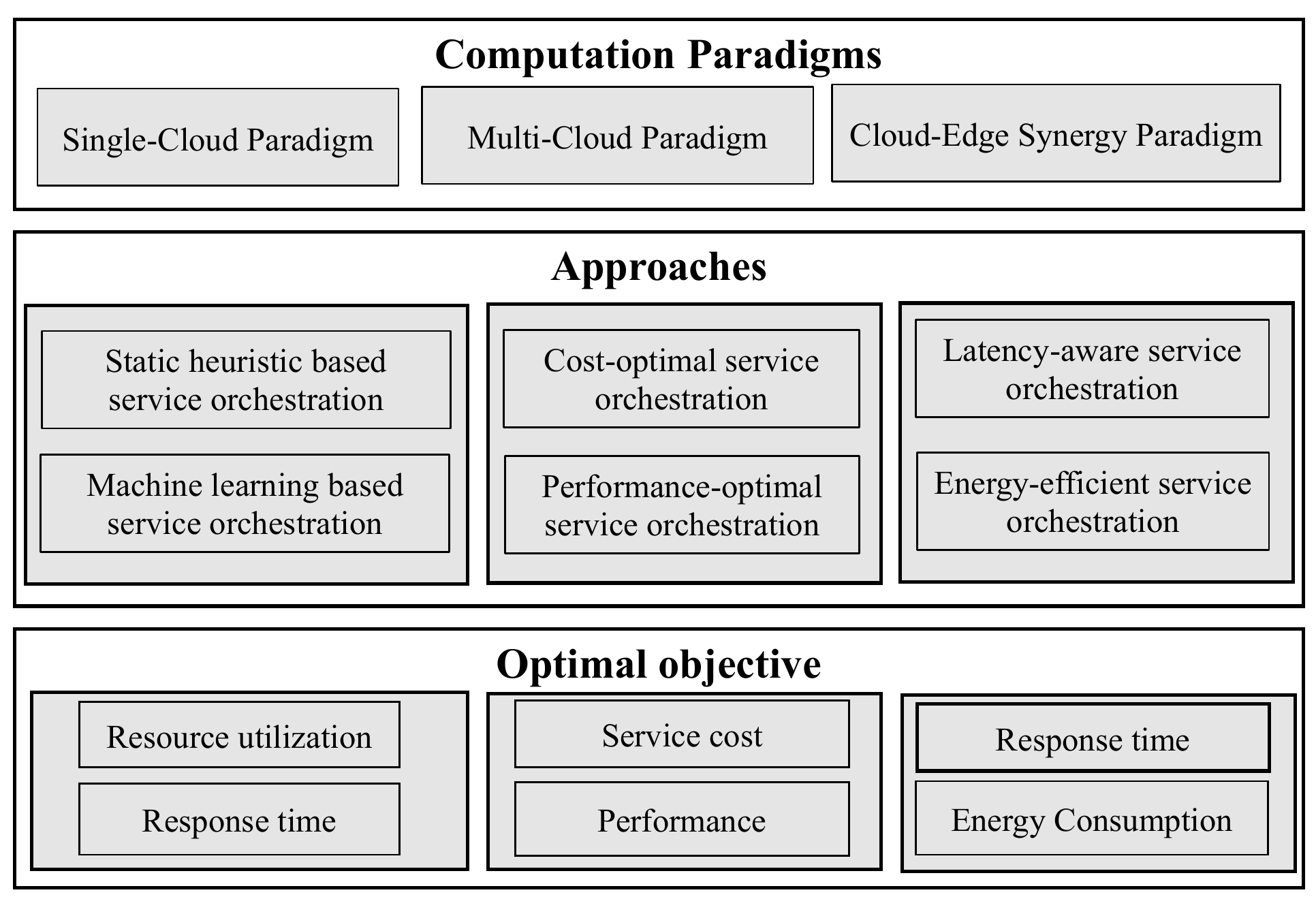}
  \caption{Service orchestration under different computation paradigms.}
  \label{computationParadigms}
\end{figure}

Existing mainstream computing paradigms include single-cloud, multi-cloud and cloud-edge synergy. Different computing paradigms have different characteristics, which have an important impact on service orchestration. Plenty of research studies have investigated the service orchestration under three different computing paradigms. We categorize them by the mainstream computing paradigms and overview them. Fig. \ref{computationParadigms} outlines the structure of this section.

\subsubsection{Single-Cloud}\label{s5.2.1}
The single-cloud paradigm is a new service model which delivers complex hardware and software services to external customers through the Internet \cite{hbb-chen2020woa}. In order to improve the utilization of cloud resource and reduce the response time of cloud service, the efficient cloud service orchestration is key. Currently, a large number of approaches are proposed to handle service orchestration \cite{hbb-wang2020elastic, hbb-zhao2018locality, hbb-zhong2020heterogeneous, hbb-carvalho2021qoe, hbb-bulej2021self, hbb-bestari2020dynamic, hbb-ambrosino2020container, hbb-chung2018stratus, hbb-zhong2020cost, hbb-zhao2022learning, hbb-zhao2022learning}. These approaches can be classified three types: static heuristic based service orchestration \cite{hbb-wang2020elastic, hbb-zhao2018locality}, machine learning based service orchestration \cite{hbb-zhong2020heterogeneous, hbb-carvalho2021qoe, hbb-bulej2021self}. For static heuristic based service orchestration, various heuristic algorithms including bin-packing algorithms, genetic algorithm, particle swarm optimization, etc., are adopted to orchestrate service in certain workload scenarios at a limited scale. For example, in \cite{hbb-wang2020elastic}, the authors formulate the placement of containers to be a variable-sized bin packing problem. To address this problem, an elastic scheduling algorithm for microservices in clouds is proposed. Its main goal is to minimize the cost of virtual machines while meeting deadline constraints. Analogously, in \cite{hbb-zhao2018locality}, the authors design some heuristic algorithms to schedule container cloud services to implement load balance and improve application performance. However, the performance of static heuristic algorithms could dramatically degrade when the system scales up. They cannot cope with the increasingly diverse and dynamic workloads and environments. To address this problem, machine learning algorithms including reinforcement learning, K-means, recurrent Neural Network, etc., are accordingly employed to orchestrate service. For example, in \cite{hbb-zhong2020heterogeneous}, a containerized task scheduler employs K-means++ algorithms to characterize the workload features and identify their behavior, and accurately co-locate heterogeneous workloads in an interference-aware manner. This scheme greatly improves resource utilization and reduces the rescheduling rate. Moreover, in \cite{hbb-carvalho2021qoe}, a Kubernetes scheduler extension which adopt machine learning algorithm to predict Quality of Experience (QoE) and schedule the resource based on the predicted QoE, is designed to improve the average QoE and eliminate over-provisioning altogether in the cloud. Also, in \cite{hbb-bulej2021self}, a self-adaptive Kubernetes scheduler (re-)deploys these time-sensitive applications by predicting their required resource in the cloud. These machine learning based service orchestration schemes can build certain machine learning model for diverse and dynamic workloads and environments and predict multi-dimensional performance metrics. These schemes could further improve the quality of resource provisioning decisions in response to the changing workloads under complex environments.


\subsubsection{Multi-Cloud}\label{s5.2.2}
With the surge of cloud workloads, single-cloud paradigm cannot meet with their various requirements, such as resource requirement, cost requirement, reliability requirement and so on. Therefore, multi-cloud paradigm is proposed. The multi-cloud paradigm enables resources among different clouds to be shared to cope with a burst of incoming tasks. In addition, the multi-cloud paradigm can efficiently improve service reliability and reduce service cost. Although benefiting from these advantages of the multi-cloud paradigm, the heterogeneity of the underlying resources and services for different cloud systems brings some new challenges to service orchestration in multi-cloud paradigm. To cope with these new challenges, some related research works about service orchestration in multi-cloud paradigm \cite{hbb-rossi2019elastic, hbb-das2020performance, hbb-akhtar2020cose, hbb-aldwyan2021elastic, hbb-shi2021location, hbb-shi2020location} are conducted. Their main optimization objectives are service cost and service performance. According to these two optimization objectives, these research works can be divided into two types: cost-optimal service orchestration and performance-optimal service orchestration. For cost-optimal service orchestration, the authors in \cite{hbb-rossi2019elastic, hbb-das2020performance, hbb-akhtar2020cose} evaluate and select diverse cloud services according to their pricing models and computing capacity, and design various service orchestration strategies to minimize the financial costs. For performance-optimal service orchestration, the authors in \cite{hbb-aldwyan2021elastic}, adopt a meta-heuristic algorithm to continuously make elastic container deployment plans in geographically distributed clouds, aim to maintain performance while minimizing the operating costs. Also, the authors in \cite{hbb-shi2020location}, propose a hybrid GA-based approach to deploy a new type of composite application in multi-cloud. Its main goal is to optimize the performance and control the budget. However, these heuristic algorithms rely on the prior knowledge of the system, and cannot cope with the high variable workloads. Thus, the authors in \cite{hbb-shi2021location} turn to learning-based method. They adopt a deep reinforcement learning to dispatch the new arriving requests for applications in multi-cloud, the goal of which is to minimize the network latency and satisfy the budget satisfaction. 

\subsubsection{Cloud-Edge Synergy}\label{s5.2.3}
With the explosive growth of data generated by the terminal devices of IoT, transmitting these massive data to remote cloud to process commonly leads to significant propagation delays, bandwidth and energy consumption. It drives the centralized cloud to sink their computation and storage resources down to the network edge to process data, which is called the cloud-edge synergy paradigm. The cloud-edge synergy paradigm has the characteristics of resource heterogeneity, device mobility and connection uncertainty. These characteristics bring some new challenges to the service orchestration in cloud-edge synergy paradigm. There are a plenty of researcher studies to investigate these challenges \cite{hbb-sami2020fscaler, hbb-chhikara2020efficient, hbb-yan2021hansel, hbb-das2020performance, hbb-han2021tailored, hbb-marchese2022network, hbb-wang2019delay, hbb-pusztai2022polaris, hbb-pusztai2021pogonip, hbb-nastic2021polaris}. Their optimization objectives mainly include response time and energy consumption. Base on their optimization objectives, we classify these studies into two types: latency-aware service orchestration and energy-efficient service orchestration. For latency-aware service orchestration, the authors in \cite{hbb-han2021tailored, hbb-wang2019delay, hbb-sami2020fscaler, hbb-tang2018migration, hbb-yan2021hansel, hbb-das2020performance}, adopt Markov decision process, reinforcement learning, deep reinforcement learning and heuristic methods to offload the containerized applications in cloud-edge synergy paradigm. Their main goals is to optimize latency. Specifically, in \cite{hbb-han2021tailored}, a learning-based scheduling framework for edge-cloud systems is designed to dispatch service request and orchestrate multiple microservices instances, the goal of which is to improve the long-term system throughput rate. Analogously, in \cite{hbb-marchese2022network}, a network-aware scheduler plugin is designed to place containerized applications on distributed cloud-edge clusters. The placement strategy of these applications considers both current network conditions and communication requirements between microservices, which is suitable for the placement of time critical applications. For energy-efficient service orchestration, the authors in \cite{hbb-chhikara2020efficient} adopt best-fit algorithms to place the containers, aiming to reduce the energy consumption. Also, in \cite{hbb-kaur2019keids}, a competent controller is presented to schedule containerized applications in edge-cloud system, aiming at minimizing the interference and the energy consumption.


\section{Service Operate}\label{s6}
Service operation, which encompasses load balancing, service migration, and resource auto-scaling, is crucial for maintaining a high-performing and efficient system infrastructure. By integrating load balancing, service migration, and resource auto-scaling in the operational state, we can enable robust and efficient service management dynamically and at scale.
Representative works are listed in Table \ref{service_operate_tbl} for summarization.

\begin{table*}[]
    \begin{center}
  \caption{\label{service_operate_tbl}Representative works in service operation.}   
  \begin{tabular}{c|ccccc|cccc|c}
  \toprule
  \multirow{2}{*}{\textsc{Work}}                                  & \multicolumn{5}{c|}{\textsc{Metric}}                                                                                                                                                                                                   & \multicolumn{4}{c|}{\textsc{Resource}}                                                                                                                                                & \multirow{2}{*}{\textsc{Algorithm}} \\
                                                                        & QoS                                            & Throughput                                     & Makespan                                       & Response time                                  & Revenue                   & CPU                                            & RAM                                            & B.W.                                           & Storage                   &                            \\ \midrule
  \cite{zc-miao2021discrete}                           & \multicolumn{1}{c|}{}                          & \multicolumn{1}{c|}{}                          & \multicolumn{1}{c|}{\checkmark} & \multicolumn{1}{c|}{}                          &                           & \multicolumn{1}{c|}{\checkmark} & \multicolumn{1}{c|}{}                          & \multicolumn{1}{c|}{\checkmark} &                           & Heuristic                  \\
  \cite{hbb-zhao2018locality}                          & \multicolumn{1}{c|}{}                          & \multicolumn{1}{c|}{\checkmark} & \multicolumn{1}{c|}{}                          & \multicolumn{1}{c|}{}                          &                           & \multicolumn{1}{c|}{\checkmark} & \multicolumn{1}{c|}{\checkmark} & \multicolumn{1}{c|}{\checkmark} & \checkmark & Heuristic                  \\
  \cite{zc-lu2022game}                                 & \multicolumn{1}{c|}{}                          & \multicolumn{1}{c|}{}                          & \multicolumn{1}{c|}{}                          & \multicolumn{1}{c|}{}                          & \checkmark & \multicolumn{1}{c|}{\checkmark} & \multicolumn{1}{c|}{}                          & \multicolumn{1}{c|}{}                          & \checkmark & Game theory                \\
  \cite{zc-ebadifard2020dynamic, zc-kumar2018deadline} & \multicolumn{1}{c|}{}                          & \multicolumn{1}{c|}{}                          & \multicolumn{1}{c|}{\checkmark} & \multicolumn{1}{c|}{}                          &                           & \multicolumn{1}{c|}{\checkmark} & \multicolumn{1}{c|}{}                          & \multicolumn{1}{c|}{}                          &                           & Heuristic                  \\
  \cite{zc-yu2019load}                                 & \multicolumn{1}{c|}{\checkmark} & \multicolumn{1}{c|}{\checkmark} & \multicolumn{1}{c|}{}                          & \multicolumn{1}{c|}{}                          &                           & \multicolumn{1}{c|}{\checkmark} & \multicolumn{1}{c|}{}                          & \multicolumn{1}{c|}{}                          &                           & LP                         \\
  \cite{zc-huang2020consistent}                        & \multicolumn{1}{c|}{\checkmark} & \multicolumn{1}{c|}{}                          & \multicolumn{1}{c|}{}                          & \multicolumn{1}{c|}{}                          &                           & \multicolumn{1}{c|}{}                          & \multicolumn{1}{c|}{}                          & \multicolumn{1}{c|}{\checkmark} & \checkmark & Heuristic                  \\
  \cite{zc-wang2021robust}                             & \multicolumn{1}{c|}{\checkmark} & \multicolumn{1}{c|}{}                          & \multicolumn{1}{c|}{}                          & \multicolumn{1}{c|}{}                          &                           & \multicolumn{1}{c|}{\checkmark} & \multicolumn{1}{c|}{}                          & \multicolumn{1}{c|}{\checkmark} &                           & Approx. algorithm    \\
  \cite{zc-gutierrez2015agent}                         & \multicolumn{1}{c|}{}                          & \multicolumn{1}{c|}{}                          & \multicolumn{1}{c|}{}                          & \multicolumn{1}{c|}{\checkmark} &                           & \multicolumn{1}{c|}{\checkmark} & \multicolumn{1}{c|}{\checkmark} & \multicolumn{1}{c|}{}                          &                           & Heuristic                  \\
  \cite{zc-menon2013distributed}                       & \multicolumn{1}{c|}{\checkmark} & \multicolumn{1}{c|}{}                          & \multicolumn{1}{c|}{}                          & \multicolumn{1}{c|}{}                          &                           & \multicolumn{1}{c|}{\checkmark} & \multicolumn{1}{c|}{}                          & \multicolumn{1}{c|}{}                          &                           & Epidemic                   \\
  \cite{zc-xu2022game}                                 & \multicolumn{1}{c|}{\checkmark} & \multicolumn{1}{c|}{}                          & \multicolumn{1}{c|}{}                          & \multicolumn{1}{c|}{}                          &                           & \multicolumn{1}{c|}{\checkmark} & \multicolumn{1}{c|}{}                          & \multicolumn{1}{c|}{\checkmark} &                           & Game theory                \\
  \cite{zc-yao2022multi}                               & \multicolumn{1}{c|}{}                          & \multicolumn{1}{c|}{}                          & \multicolumn{1}{c|}{\checkmark} & \multicolumn{1}{c|}{}                          &                           & \multicolumn{1}{c|}{\checkmark} & \multicolumn{1}{c|}{}                          & \multicolumn{1}{c|}{}                          &                           & MARL                       \\
  \cite{zc-asghari2022combined}                        & \multicolumn{1}{c|}{}                          & \multicolumn{1}{c|}{}                          & \multicolumn{1}{c|}{}                          & \multicolumn{1}{c|}{\checkmark} &                           & \multicolumn{1}{c|}{\checkmark} & \multicolumn{1}{c|}{}                          & \multicolumn{1}{c|}{}                          &                           & DQN                        \\
  \cite{zc-houidi2022multi}                            & \multicolumn{1}{c|}{}                          & \multicolumn{1}{c|}{\checkmark} & \multicolumn{1}{c|}{}                          & \multicolumn{1}{c|}{\checkmark} &                           & \multicolumn{1}{c|}{\checkmark} & \multicolumn{1}{c|}{}                          & \multicolumn{1}{c|}{\checkmark} &                           & MARL                       \\
  \cite{zc-shribman2012pre}                            & \multicolumn{1}{c|}{}                          & \multicolumn{1}{c|}{}                          & \multicolumn{1}{c|}{}                          & \multicolumn{1}{c|}{\checkmark} &                           & \multicolumn{1}{c|}{\checkmark} & \multicolumn{1}{c|}{\checkmark} & \multicolumn{1}{c|}{}                          &                           & Heuristic                  \\
  \cite{zc-fernando2019live, zc-chou2019optimizing}    & \multicolumn{1}{c|}{}                          & \multicolumn{1}{c|}{}                          & \multicolumn{1}{c|}{}                          & \multicolumn{1}{c|}{\checkmark} &                           & \multicolumn{1}{c|}{\checkmark} & \multicolumn{1}{c|}{\checkmark} & \multicolumn{1}{c|}{}                          &                           & -                          \\
  \cite{zc-jo2017machine}                              & \multicolumn{1}{c|}{}                          & \multicolumn{1}{c|}{}                          & \multicolumn{1}{c|}{}                          & \multicolumn{1}{c|}{\checkmark} &                           & \multicolumn{1}{c|}{\checkmark} & \multicolumn{1}{c|}{\checkmark} & \multicolumn{1}{c|}{\checkmark} &                           & Linear regression          \\
  \cite{zc-khai2017multi}                              & \multicolumn{1}{c|}{\checkmark} & \multicolumn{1}{c|}{}                          & \multicolumn{1}{c|}{}                          & \multicolumn{1}{c|}{}                          & \checkmark & \multicolumn{1}{c|}{\checkmark} & \multicolumn{1}{c|}{}                          & \multicolumn{1}{c|}{\checkmark} &                           & MIP                        \\
  \cite{zc-ruprecht2018vm}                             & \multicolumn{1}{c|}{}                          & \multicolumn{1}{c|}{}                          & \multicolumn{1}{c|}{}                          & \multicolumn{1}{c|}{}                          & \checkmark & \multicolumn{1}{c|}{\checkmark} & \multicolumn{1}{c|}{}                          & \multicolumn{1}{c|}{\checkmark} &                           & -                          \\
  \cite{zc-li2017bac}                                  & \multicolumn{1}{c|}{}                          & \multicolumn{1}{c|}{}                          & \multicolumn{1}{c|}{}                          & \multicolumn{1}{c|}{\checkmark} &                           & \multicolumn{1}{c|}{\checkmark} & \multicolumn{1}{c|}{}                          & \multicolumn{1}{c|}{\checkmark} &                           & -                          \\
  \cite{zc-le2019experiences}                          & \multicolumn{1}{c|}{}                          & \multicolumn{1}{c|}{\checkmark} & \multicolumn{1}{c|}{}                          & \multicolumn{1}{c|}{\checkmark} &                           & \multicolumn{1}{c|}{\checkmark} & \multicolumn{1}{c|}{}                          & \multicolumn{1}{c|}{\checkmark} &                           & -                          \\
  \cite{zc-basu2019learn}                              & \multicolumn{1}{c|}{\checkmark} & \multicolumn{1}{c|}{}                          & \multicolumn{1}{c|}{}                          & \multicolumn{1}{c|}{\checkmark} & \checkmark & \multicolumn{1}{c|}{\checkmark} & \multicolumn{1}{c|}{}                          & \multicolumn{1}{c|}{}                          &                           & RL                         \\
  \cite{zc-benjaponpitak2020enabling}                  & \multicolumn{1}{c|}{}                          & \multicolumn{1}{c|}{\checkmark} & \multicolumn{1}{c|}{}                          & \multicolumn{1}{c|}{\checkmark} &                           & \multicolumn{1}{c|}{\checkmark} & \multicolumn{1}{c|}{\checkmark} & \multicolumn{1}{c|}{\checkmark} &                           & -                          \\
  \cite{zc-sinha2019mwarp}                             & \multicolumn{1}{c|}{}                          & \multicolumn{1}{c|}{}                          & \multicolumn{1}{c|}{}                          & \multicolumn{1}{c|}{\checkmark} &                           & \multicolumn{1}{c|}{}                          & \multicolumn{1}{c|}{\checkmark} & \multicolumn{1}{c|}{\checkmark} &                           & -                          \\
  \cite{zc-xu2020sledge}                               & \multicolumn{1}{c|}{\checkmark} & \multicolumn{1}{c|}{}                          & \multicolumn{1}{c|}{}                          & \multicolumn{1}{c|}{\checkmark} &                           & \multicolumn{1}{c|}{}                          & \multicolumn{1}{c|}{\checkmark} & \multicolumn{1}{c|}{\checkmark} &                           & -                          \\
  \cite{zc-torre2019towards}                           & \multicolumn{1}{c|}{}                          & \multicolumn{1}{c|}{}                          & \multicolumn{1}{c|}{}                          & \multicolumn{1}{c|}{\checkmark} &                           & \multicolumn{1}{c|}{\checkmark} & \multicolumn{1}{c|}{}                          & \multicolumn{1}{c|}{\checkmark} &                           & -                          \\
  \cite{zc-romero2021infaas, zc-zhang2019mark}         & \multicolumn{1}{c|}{}                          & \multicolumn{1}{c|}{\checkmark} & \multicolumn{1}{c|}{}                          & \multicolumn{1}{c|}{\checkmark} & \checkmark & \multicolumn{1}{c|}{\checkmark} & \multicolumn{1}{c|}{}                          & \multicolumn{1}{c|}{\checkmark} &                           & Heuristic                  \\
  \cite{zc-shillaker2020faasm}                         & \multicolumn{1}{c|}{}                          & \multicolumn{1}{c|}{\checkmark} & \multicolumn{1}{c|}{}                          & \multicolumn{1}{c|}{\checkmark} &                           & \multicolumn{1}{c|}{\checkmark} & \multicolumn{1}{c|}{\checkmark} & \multicolumn{1}{c|}{\checkmark} &                           & -                          \\
  \cite{zc-akhtar2018sec}                              & \multicolumn{1}{c|}{}                          & \multicolumn{1}{c|}{\checkmark} & \multicolumn{1}{c|}{}                          & \multicolumn{1}{c|}{\checkmark} &                           & \multicolumn{1}{c|}{\checkmark} & \multicolumn{1}{c|}{}                          & \multicolumn{1}{c|}{}                          &                           & PID                        \\
  \cite{zc-russo2021mead}                              & \multicolumn{1}{c|}{}                          & \multicolumn{1}{c|}{}                          & \multicolumn{1}{c|}{}                          & \multicolumn{1}{c|}{\checkmark} &                           & \multicolumn{1}{c|}{\checkmark} & \multicolumn{1}{c|}{}                          & \multicolumn{1}{c|}{}                          &                           & MAP                        \\
  \cite{zc-lakew2017kpi}                               & \multicolumn{1}{c|}{}                          & \multicolumn{1}{c|}{\checkmark} & \multicolumn{1}{c|}{}                          & \multicolumn{1}{c|}{\checkmark} &                           & \multicolumn{1}{c|}{\checkmark} & \multicolumn{1}{c|}{\checkmark} & \multicolumn{1}{c|}{}                          &                           & MPC                        \\
  \cite{zc-sfakianakis2022latest}                      & \multicolumn{1}{c|}{}                          & \multicolumn{1}{c|}{}                          & \multicolumn{1}{c|}{}                          & \multicolumn{1}{c|}{\checkmark} &                           & \multicolumn{1}{c|}{\checkmark} & \multicolumn{1}{c|}{\checkmark} & \multicolumn{1}{c|}{}                          &                           & PID                        \\
  \cite{zc-tesfatsion2017optibook}                     & \multicolumn{1}{c|}{\checkmark} & \multicolumn{1}{c|}{}                          & \multicolumn{1}{c|}{}                          & \multicolumn{1}{c|}{\checkmark} &                           & \multicolumn{1}{c|}{\checkmark} & \multicolumn{1}{c|}{}                          & \multicolumn{1}{c|}{}                          &                           & Fuzzy logic                \\
  \cite{zc-fei2018adaptive}                            & \multicolumn{1}{c|}{\checkmark} & \multicolumn{1}{c|}{\checkmark} & \multicolumn{1}{c|}{}                          & \multicolumn{1}{c|}{}                          & \checkmark & \multicolumn{1}{c|}{\checkmark} & \multicolumn{1}{c|}{}                          & \multicolumn{1}{c|}{\checkmark} &                           & FTRL                       \\
  \cite{zc-avgeris2019adaptive}                        & \multicolumn{1}{c|}{\checkmark} & \multicolumn{1}{c|}{}                          & \multicolumn{1}{c|}{}                          & \multicolumn{1}{c|}{\checkmark} &                           & \multicolumn{1}{c|}{\checkmark} & \multicolumn{1}{c|}{\checkmark} & \multicolumn{1}{c|}{}                          &                           & ILP                        \\
  \cite{zc-mahmud2018latency}                          & \multicolumn{1}{c|}{\checkmark} & \multicolumn{1}{c|}{}                          & \multicolumn{1}{c|}{}                          & \multicolumn{1}{c|}{\checkmark} &                           & \multicolumn{1}{c|}{\checkmark} & \multicolumn{1}{c|}{\checkmark} & \multicolumn{1}{c|}{\checkmark} &                           & Heuristic                  \\
  \cite{zc-schuler2021ai}                              & \multicolumn{1}{c|}{}                          & \multicolumn{1}{c|}{\checkmark} & \multicolumn{1}{c|}{}                          & \multicolumn{1}{c|}{\checkmark} &                           & \multicolumn{1}{c|}{\checkmark} & \multicolumn{1}{c|}{}                          & \multicolumn{1}{c|}{\checkmark} &                           & RL                         \\ \bottomrule
  \end{tabular}
  \end{center}
  \end{table*}

\subsection{Load Balancing}\label{s6.1}
Load balancing is a technique used to address the problem of workload imbalance across multiple containers. It enables optimal utilization of resources, improves throughput, and reduces response time and makespan. In cloud-native environments, the primary goal of load balancing is to prevent overloading of a single container or cluster while keeping other containers idle.
Cloud-native applications with high throughput and parallel computing architectures require effective load balancing techniques. One such technique involves redistributing heavy workloads from a single virtual server to multiple virtual servers, ensuring optimal resource utilization. In Sec. \ref{sec:lb-algo}, we will provide a comprehensive analysis of load-balancing algorithms.
Sec. \ref{sec:lb-tech} introduces the current techniques for implementing load balancing in cloud-native. 

\subsubsection{Algorithm Design and Analysis}\label{sec:lb-algo}
\begin{figure}[htbp]
  \centering
  \includegraphics[width=3in]{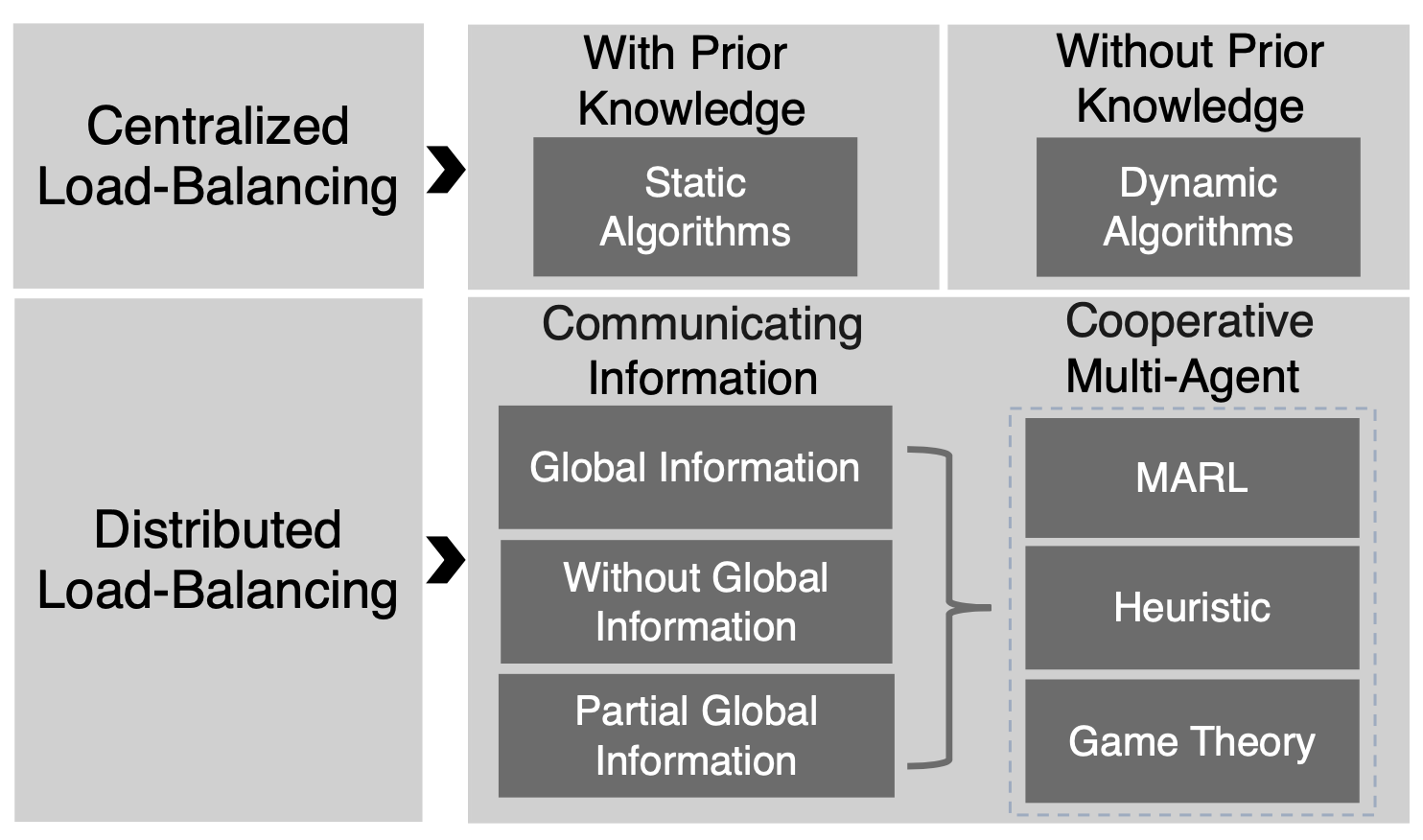}
  \caption{Algorithms for load-balancing.}
  \label{load-balancing}
\end{figure}

Load balancing can be divided into two categories: centralized and distributed, as illustrated in Fig. \ref{load-balancing}. Centralized load balancing can further be classified into static and dynamic algorithms based on whether the algorithm incorporates prior knowledge of the system. On the other hand, distributed load balancing employs multi-agent algorithms that offer greater flexibility and scalability than centralized algorithms.

\textbf{Static algorithms}.
In cloud-native environments, static load-balancing algorithms are widely adopted due to their ability to leverage prior knowledge of system states. These algorithms, such as Round-robin, FIFO, MIN-MIN, and others, do not require detailed information about the current workload running in the system. Instead, they rely on factors such as CPU usage, memory usage, storage availability, or network bandwidth utilization. Implementing static algorithms in a cloud-native system is generally straightforward. Heuristic algorithms play a significant role in developing load-balancing strategies for cloud systems under fixed conditions. For example, Zhang et al. \cite{zc-miao2021discrete} propose a static algorithm called APDPSO that utilizes a particle swarm optimization method. They treat the allocation of suitable hosts as a discrete optimization problem. Similarly, Zhao et al. \cite{hbb-zhao2018locality} propose a load-balancing solution based on service performance. They employ a statistical method to optimize the collaboration problem heuristically.

While static load-balancing algorithms are efficient and easy to implement, it is essential to consider server performance and continuously monitor the current load status to prevent exacerbating load imbalances during long-term execution. Regular maintenance and adjustment of the load-balancing algorithm are necessary to ensure optimal performance and resource utilization.

\textbf{Dynamic algorithms}.
Dynamic load-balancing algorithms offer superior performance in adaptively executing load balancing, especially when dealing with sudden changes in workload. They can effectively operate on a cloud-native platform without prior knowledge of the system or workload. These algorithms adjust the allocation of resources dynamically based on real-time information, such as server availability and network bandwidth utilization, ensuring optimal resource utilization and reduced response times.

Load balancing involves transferring tasks to appropriate servers to alleviate the burden on overloaded servers. One crucial aspect is designing an efficient scheduling strategy to minimize the average load across all servers and reduce the makespan of the system. Hongli et al. \cite{zc-lu2022game} propose a theoretical game algorithm that balances tasks by offloading them to edge servers while ensuring Service Level Objectives (SLOs). Another approach to dynamic load balancing revolves around resource allocation. Fatemeh et al. \cite{zc-ebadifard2020dynamic} design a heuristic algorithm-based optimization algorithm that evaluates overloaded, loaded, and balanced virtual machines to achieve load balancing. However, these efficiency-focused load-balancing algorithms may not be suitable for large-scale virtual machine environments.

To meet the requirements of reliability and elasticity in large-scale systems, Mohit et al. \cite{zc-kumar2018deadline} propose a dynamic scheduling algorithm based on the last optimal $k$-interval virtual machines strategy that balances workload through resource provisioning and de-provisioning methods. This algorithm effectively balances the workload of virtual machines in a cloud environment through resource provisioning and de-provisioning methods. 
Regarding load balancing among microservices, Ruozhou et al. \cite{zc-yu2019load} introduce a graph-based model to analyze dependencies among microservices and adopted a polynomial approximation method to solve the QoS-aware load-balancing optimization problem.
Network traffic is another critical metric for monitoring the state of cloud-native systems, often triggering load-balancing operations. Lemei et al. \cite{zc-huang2020consistent} address service unreliability and dynamic network traffic challenges by designing a load-balancing strategy based on traffic allocation consistency and DNS granularity, aiming to achieve an approximate solution to the QoS optimization problem. To efficiently route network traffic, Jingzhou et al. \cite{zc-wang2021robust} propose an approximation algorithm with a polynomial-time complexity that follows a two-step process to implement service deployment.

\textbf{Multi-agent algorithms}.
In large-scale clusters, centralized load-balancing solutions can become time-consuming due to the reliance on a single machine for decision-making. These solutions gather system information from the involved servers, which introduces delays in the decision-making process. On the other hand, distributed load-balancing schemes offer advantages in terms of scalability and flexibility, particularly in cloud-native environments. Imbalanced workloads among heterogeneous servers in the cloud can result in performance degradation within the cloud platform. To address this challenge, Gutierrez et al. \cite{zc-gutierrez2015agent} design a distributed approach that focuses on migrating virtual machines (VMs) to achieve load balancing. Their approach outperforms the centralized load-balancing method. However, it should be noted that collecting global information for load-balancing decisions in each agent can still be time-consuming. To tackle this issue, Harshitha et al. \cite{zc-menon2013distributed} propose a distributed load-balancing scheme that leverages partial information about the global state of the cloud system. Their scheme involves two steps: global information propagation and workload transfer. By utilizing partial information, the load-balancing process can be expedited while still achieving effective load distribution. Another proposed scheme, F-TORA, by Xu et al. \cite{zc-xu2022game}, focuses on task load balancing. It utilizes fuzzy neural networks and game theory to optimize task allocation and resource utilization. F-TORA aims to ensure timely and high-quality services by intelligently distributing tasks among available resources. These distributed load-balancing schemes offer advantages over centralized solutions in cloud-native environments. They provide scalability, flexibility, and improved performance by efficiently distributing workloads across servers or VMs. However, it's important to consider the specific requirements and characteristics of the system before choosing the most suitable load-balancing scheme.

The advent of intelligent algorithms, such as reinforcement learning, has opened up new possibilities for distributed load balancing. Reinforcement learning techniques, including Q-learning and multi-agent reinforcement learning (MARL), have gained popularity in this domain.
Zhiyuan et al. \cite{zc-yao2022multi} propose a MARL framework that specifically addresses the dynamics of arrival workload. This framework overcomes the limitations of independent and selfish algorithms commonly used in load-balancing schemes. By leveraging MARL, the proposed approach enables agents to collaborate and make coordinated decisions, leading to more effective load balancing in dynamic workload scenarios.
Ali et al. \cite{zc-asghari2022combined} design a multi-agent deep Q-network with coral reefs optimization (MDQ-CR) to minimize the energy consumption of cloud computing. This approach combines the power of deep Q-networks, a variant of reinforcement learning, with coral reefs optimization, a nature-inspired optimization algorithm. The combination of these techniques enables efficient load balancing while considering energy consumption as a critical factor.
Omar et al. \cite{zc-houidi2022multi} utilize a graph neural network (GNN)-based method to model the network as a graph and apply MARL techniques to tackle the load-balancing problem while scheduling traffic flow. By representing the network as a graph, the authors capture the dependencies between nodes and leverage GNNs to process and aggregate information effectively. MARL techniques are then used to optimize load balancing and traffic scheduling based on the learned graph representations.
These studies highlight the application of reinforcement learning, particularly MARL, in distributed load balancing. These intelligent algorithms provide a promising avenue for addressing load-balancing challenges and optimizing various aspects such as workload dynamics, energy consumption, and traffic flow in cloud computing environments.

\subsubsection{Tools and Systems}\label{sec:lb-tech}
The most widely used load balancing techniques possess several desirable characteristics, including scalability, flexibility, low cost, simple deployment, and security. The load balancer allows the system to adapt to dynamic workloads by scaling in or out as needed. It should work seamlessly with various operating systems, cloud environments, and virtual machines and can be easily deployed. Additionally, the load balancer should provide a secure environment for the system and its users. Some popular load balancing solutions include Nginx\cite{zc-nginx}, a widely-deployed reverse proxy server, and HAProxy\cite{zc-HAProxy}, a fast and efficient reverse proxy software.
Recent advancements in load balancing technology include Maglev\cite{zc-eisenbud2016maglev}, which is able to balance sudden spikes in network traffic based on ECMP rules. Maglev is Google's production load balancer, which fully utilizes multiple networking techniques to achieve flexible and scalable load balancing. Specifically, Maglev utilizes Google's global backbone to announce IP prefixes at the same cost so that BGP routers can provide the first layer of load balancing. Then, IP packets are evenly distributed among service endpoints, providing another layer of load balancing. Since Maglev is entirely software-based, adding more load-balancing capability is simple as long as the backbone or service endpoints are not saturated.  CHEETAH\cite{zc-barbette2020high} is another load balancer that supports uniform load distribution with per-connection consistency. 

\subsection{Service Migration}
Service migration refers to moving the service application from the original clouds or machines to the destination.
The host transfers all system states, including the memory, file system, and network connectivity profiles, to the destination host, keeping conditions without changes.

\subsubsection{Algorithm Design and Analysis}
\begin{figure}[htbp]
  \centering
  \includegraphics[width=3in]{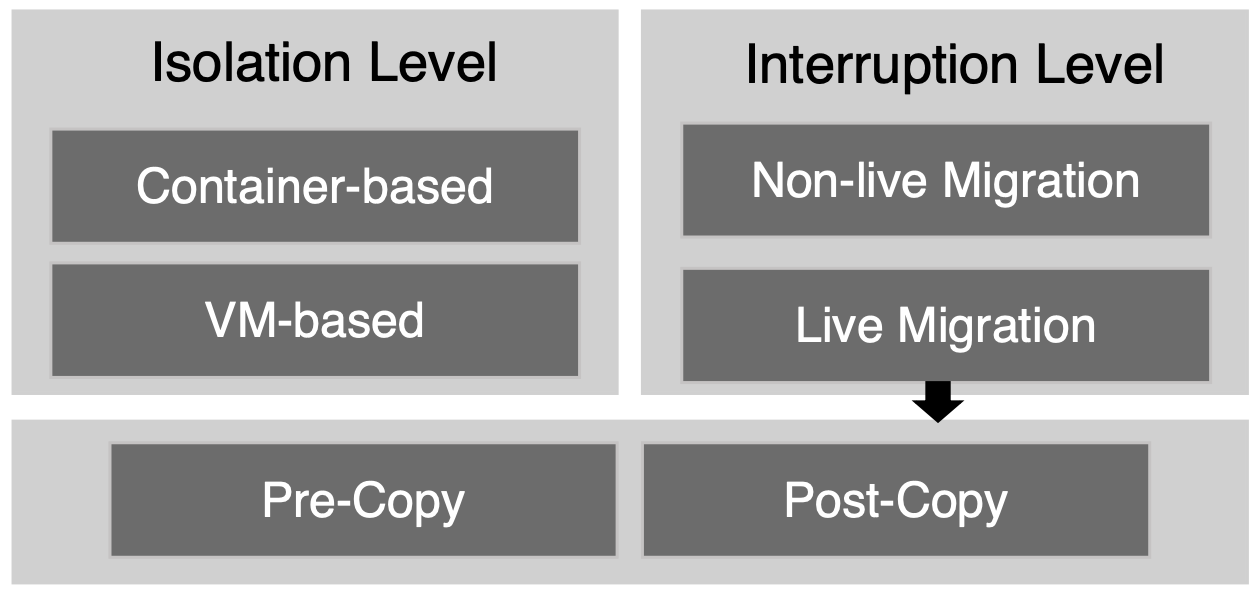}
  \caption{Classification of service migration.}
  \label{service-migration}
\end{figure}

Service migration is a critical aspect of cloud-native environments, encompassing live migration, VM-based migration, and container-based migration. Live migration offers minimal impact on running services and preserves memory data. VM-based migration focuses on optimizing the process through modeling, prediction, and analysis. Container-based migration benefits from efficient migration techniques and tools, enabling seamless migration of container-based services. Evaluating migration performance under various conditions is essential. Service migration enhances the flexibility, efficiency, and reliability of cloud-native systems.
We introduce service migration in the following aspects as shown in Fig. \ref{service-migration}.

\textbf{Live vs. non-live service migration.}
The main difference between live and non-live migration is as follows: non-live migration requires the system to be shut down during the migration process, while live migration involves migrating running systems. Non-live migration is a simpler approach but does not support the preservation of memory data, leading to memory data loss. On the other hand, live migration offers the advantage of minimal impact on running services, with low interruption time and the ability to preserve data in memory. Additionally, applications running in the system remain unaware of the migration. However, live migration can be a complex operation, and the migration process may encounter interruptions.
Despite the challenges associated with live migration, it is widely used in cloud-native environments due to the flexibility it provides through its techniques \cite{zc-elsaid2022virtual}.

Live migration predominantly employs two methods: pre-copy and post-copy \cite{zc-shribman2012pre}. The pre-copy migration algorithm involves iterative copy operations, which can sometimes result in the migration process failing to converge, leading to prolonged overall migration time. On the other hand, the post-copy migration algorithm offers shorter overall migration time. However, this approach can cause page faults during the migration process, resulting in degraded performance and reduced stability of the virtual machine \cite{zc-noshy2018optimization}.
To address the challenges of post-copy migration, fault tolerance becomes necessary after a failure during the recovery of a virtual machine. One method, called PostCopyFT, tackles this issue by utilizing an efficient reverse incremental checkpoint mechanism. This approach resolves the problem without increasing the total migration time \cite{zc-fernando2019live}.
Additionally, an optimized post-copy mechanism based on Fabric-Attached Memories (FAMs) has been proposed. This mechanism, which is FAM-aware and employs system-level checkpointing, reduces both the migration time and the system's busy time \cite{zc-chou2019optimizing}.
These advancements in live migration techniques aim to improve the efficiency, reliability, and performance of the migration process. The choice between pre-copy and post-copy methods depends on factors such as migration time, system stability, fault tolerance requirements, and the impact on the virtual machine's performance.

\textbf{VM-based vs. container-based service migration.}
Existing works have optimized the VM-based migration process from various perspectives including modeling \cite{zc-jo2017machine, zc-khai2017multi} , prediction \cite{zc-jo2017machine, zc-ruprecht2018vm} , latency\cite{zc-li2017bac, zc-le2019experiences}, energy consumption \cite{zc-basu2019learn}, etc.
The dynamics of workloads make live migration modeling challenging. Jo et al. propose a machine learning-based model to enhance the prediction accuracy, considering critical characteristics of live migration \cite{zc-jo2017machine}.
Nguyen et al. develop a two-phase migration optimization model aimed at optimizing VM movement. The first phase computes an optimal embedding strategy to reduce demands on other virtual networks, while the second phase executes the migration using this strategy \cite{zc-khai2017multi}.
Maintaining uninterrupted uptime is crucial for live migration, particularly in large-scale systems with frequent infrastructure changes. Adam et al. propose a live VM migration scheme that minimizes the impact on users while addressing version updates and security concerns \cite{zc-ruprecht2018vm}.
Bandwidth-Aware Compression (BAC) focuses on the trade-off between VM compression and transmission during migration \cite{zc-li2017bac}. The utilization of multi-page compression techniques enables an efficient migration scheme, reducing total migration latency while maintaining performance comparable to benchmarks.
To enhance performance during live migration, Franck et al. propose a new Multi-Path TCP method over WAN, which significantly decreases round-trip latency and improves responsiveness and user engagement \cite{zc-le2019experiences}.
For energy consumption reduction and resource allocation in cloud-native environments, Basu et al. adopt a reinforcement learning algorithm to make optimal decisions regarding virtual machine migration \cite{zc-basu2019learn}.
These research efforts aim to address various challenges and optimize live migration processes in terms of prediction accuracy, network optimization, uninterrupted uptime, bandwidth management, performance improvement, and resource allocation.

Container-based migration has gained popularity in cloud-native environments compared to VM-based migration \cite{zc-ma2018efficient, zc-benjaponpitak2020enabling, zc-sinha2019mwarp, zc-xu2020sledge, zc-torre2019towards}. Cloud-native platforms like K8s and Docker Swarm offer efficient handoff capabilities during container migration.
To reduce handoff latency during migration, Lele et al. propose a framework that enables mobile users to offload their tasks to edge servers through seamless migration of container-based services \cite{zc-ma2018efficient}.
In order to provide users with the freedom to choose cloud-native platforms, Thad et al. introduce a tool called CloudHopper, which facilitates the movement of containers between different platforms \cite{zc-benjaponpitak2020enabling}.
Live migration is widely utilized in cloud-native platforms, but the cost of copying numerous memory pages from a source to a destination server can be high. To tackle this challenge, Piush et al. present mWarp, a live container migration tool that efficiently remaps the physical memory of containers \cite{zc-sinha2019mwarp}.
Bo proposes an efficient live migration system called Sledge, which integrates images and management context to reduce migration overhead and improve quality of service (QoS) with minimal downtime. The system employs a dynamic context-loading mechanism to minimize downtime during migration \cite{zc-xu2020sledge}.
Although containers boot faster than VMs, their behavior during live migration under non-ideal conditions remains a question. Roberto et al. develop a testbed to evaluate latency and downtime during live container migration in adjusted conditions. They find that network overload significantly impacts migration performance, while stressing a container within a host has minimal effect \cite{zc-torre2019towards}.
These advancements in container-based migration address various challenges and offer solutions for efficient handoff, provider flexibility, memory optimization, migration overhead reduction, and evaluation of migration performance in different conditions.

\subsubsection{Tools and Systems}
Service migration aims to solve problems such as the upgrade during service operations, load balancing between clusters, and service deployment between cloud vendors. 
The most popular hypervisors used for migration in cloud-native are as follows.
\begin{itemize}
    \item 
       Kernel-based Virtual Machine (KVM) is a module in the Linux kernel used to virtualize physical machines. 
       It enables the host machine to turn into a hypervisor running multiple isolated virtual environments. KVM was first announced in 2006 and merged into Linux kernel releases a year later\cite{zc-KVM}. 
    \item
    Xen focuses on the virtualization technology that supports multiple cloud platforms. 
        The most significant feature of Xen is that it can support multiple guest operating systems, for instance, Linux, Windows, NetBSD, and FreeBSD, etc. 
        It allows live migration between multiple hosts seamlessly\cite{zc-Xen}.
    \item
    OpenVZ is a virtualization technology for Linux based on an operating system level. 
        It can support multiple operating systems and allow live container migration using checkpointing features with little delay\cite{zc-OpenVz}.
    \item
        Checkpoint/Restore In Userspace (CRIU) is a software tool for Linux to freeze the system states by a checkpoint technology.
        With CRIU, we can operate live migration in user space which is mainly distinctive to other migration tools.
        During live migration, CRIU can convert the frozen running applications into a collection of files and then restore them in the checkpoint frozen\cite{zc-criu}.
\end{itemize}

\subsection{Resource Auto-Scaling}
With the  \textit{pay-as-you-go} principle, a cloud vendor allows applications to dynamically acquire or release their resources on their demands. 
Thus, the application provider can leverage the \textit{auto-scaling} method to efficiently utilize the elastic feature of resources according to its budget and profit.
This section introduces \textit{auto-scaling} in three categories, i.e., vertical scaling, horizontal scaling, and hybrid auto-scaling in Sec. \ref{sec:atscl_categ}. Tools and systems are presented in Sec. \ref{sec:atscl_tech}. 

\subsubsection{Algorithm Design and Analysis}\label{sec:atscl_categ}
\begin{figure}[htbp]
  \centering
  \includegraphics[width=3in]{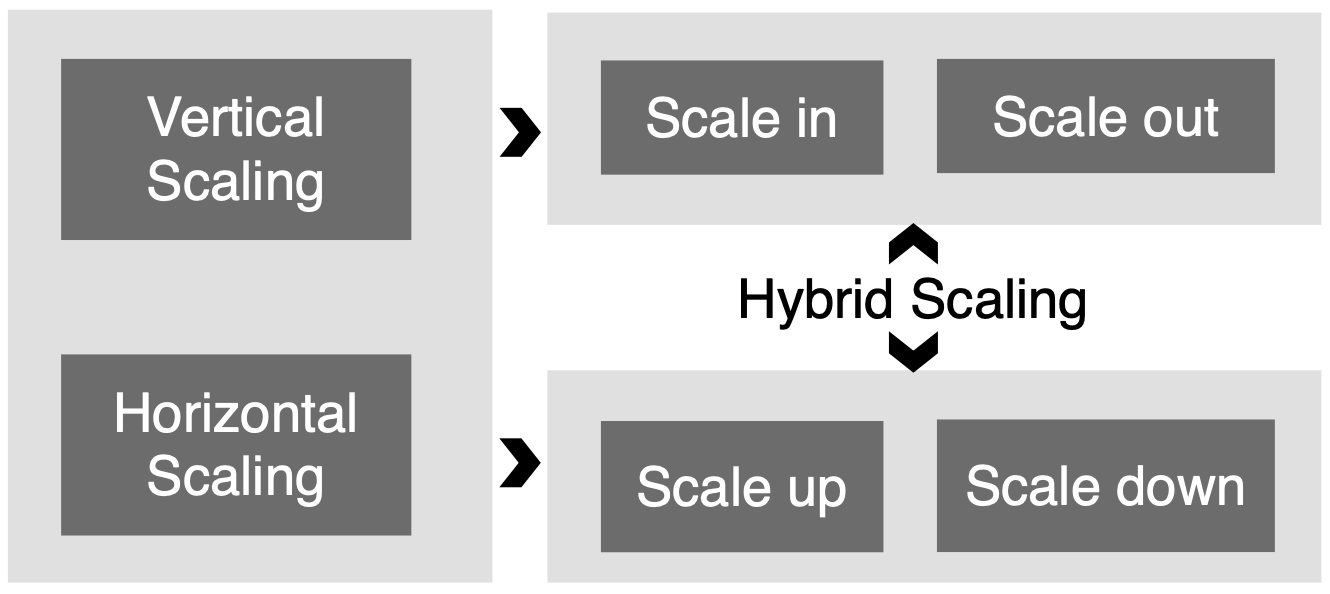}
  \caption{Techniques for Resource Auto-Scaling.}
  \label{auto-scaling}
\end{figure}

According to different policies adopted by auto-scaling, we summarise auto-scaling into three categories as shown in Fig. \ref{auto-scaling}.

\textbf{Horizontal}.
Horizontal scaling is a widely adopted approach for auto-scaling in cloud-native environments, enabling applications to dynamically adjust the number of virtual machines (VMs) to scale resources. Researchers have explored cost-efficient methods \cite{zc-romero2021infaas, zc-zhang2019mark} and fast scaling approaches \cite{zc-somma2020less, zc-shillaker2020faasm} in this domain.
Cost-effectiveness is a crucial consideration in implementing horizontal scaling in cloud-native systems. Romero et al. introduce INFaaS, which optimizes resource cost efficiency for machine learning inference applications with evolving dynamic requirements \cite{zc-romero2021infaas}. Zhang et al. propose a solution for auto-scaling in ML-as-a-Service using an LSTM network for workload prediction and a heuristic method for optimal instance provisioning decisions \cite{zc-zhang2019mark}.
The ability to respond quickly with guaranteed response times is a key requirement in cloud-native platforms. Somma et al. present a fast resource provisioning method consisting of deploying containers responsible for application services and auto-scaling resource allocations among containers \cite{zc-somma2020less}. Shillaker et al. design Faaslets, a lightweight horizontal-scaling approach for containers in clusters \cite{zc-shillaker2020faasm}.
In the context of virtualized network functions (VNFs) in cloud-native systems, auto-scaling techniques have a significant impact on on-the-fly provisioned infrastructure performance. Salhab et al. propose a framework to address resource provisioning on-demand for the 5G core network through auto-scaling of constrained resources \cite{zc-salhab2019nfv}. Akhtar et al. design a horizontal scaling manager on virtualized infrastructure to balance traffic workload for security network functions \cite{zc-akhtar2018sec}.
These research efforts contribute to addressing the challenges of cost-effectiveness, fast scaling, and resource provisioning in horizontal scaling for cloud-native environments, benefiting applications with improved efficiency, responsiveness, and performance.

\textbf{Vertical}.
Vertical scaling, which involves adjusting resources within an individual VM such as CPU, RAM, and storage, allows applications to modify their serviceability. Recent studies have focused on optimizing the cost and efficiency of vertical scaling in cloud-native environments \cite{zc-russo2021mead, zc-lakew2017kpi, zc-sfakianakis2022latest, zc-tesfatsion2017optibook}.
To achieve cost-effective resource allocation in vertical scaling, Russo et al. propose MEAD, which utilizes a prediction algorithm based on Markovian Arrival Processes to handle bursty workloads, along with an auto-scaling module for resource allocation \cite{zc-russo2021mead}. Lakew et al. address the resource allocation problem by employing a fine-grained vertical scaling technique that adapts to varying workloads in cloud-native systems \cite{zc-lakew2017kpi}.
Efficiency improvement in vertical scaling has also been a focus of research. Sfakianakis et al. introduce LatEst, a vertical scaling strategy that predicts bursts in serverless cloud systems and allocates resources efficiently within minimal time \cite{zc-sfakianakis2022latest}. Tesfatsion et al. aim to increase resource usage and reduce energy consumption through long-term optimization using vertical scaling techniques \cite{zc-tesfatsion2017optibook}.
In the context of avoiding overloaded Virtualized Network Functions (VNFs), Fei et al. propose an approximation algorithm that minimizes the prediction error caused by VNF workload, followed by the implementation of a vertical scaling technique to achieve load balancing for VNFs \cite{zc-fei2018adaptive}.
These studies contribute to the optimization of cost, efficiency, and workload management in vertical scaling, enabling applications to adapt their resource allocation dynamically within individual VMs in cloud-native environments.

\textbf{Hybrid}.
Hybrid scaling is a method commonly used in cloud-native networks that combines horizontal and vertical scaling mechanisms simultaneously. This approach leverages the advantages of both horizontal scaling and vertical scaling, making it more flexible and robust in managing resource provisioning.
To achieve efficient resource provisioning in hybrid scaling, Shahidinejad et al. utilize the Imperialist Competition Algorithm (ICA) and K-means methods to evaluate the workload from users. Based on these evaluations, they make optimal decisions using a combination of horizontal and vertical scaling techniques \cite{zc-shahidinejad2021resource}. Avgeris et al. employ a control-theoretic approach to establish a hybrid scaling method that maximizes the number of offloading requests in a cloud-native edge network, aiming for efficient resource allocation \cite{zc-avgeris2019adaptive}.
In terms of minimizing resource costs, Mahmud et al. propose a framework that integrates a latency-aware and deadline-satisfied strategy in a hybrid scaling approach. This framework optimizes the number of edge nodes required to meet application requirements while minimizing resource expenses \cite{zc-mahmud2018latency}. Schuler et al. introduce a reinforcement learning-based algorithm to minimize resource provisioning in serverless environments. By adopting a hybrid scaling method, they dynamically adjust resources to meet the dynamic demands of users while optimizing resource allocation \cite{zc-schuler2021ai}.
These studies demonstrate the benefits of hybrid scaling in cloud-native networks, allowing for efficient and cost-effective resource provisioning by combining horizontal and vertical scaling techniques.

\subsubsection{Tools and Systems
}\label{sec:atscl_tech}
The typical tools, plugins, and systems that are used for auto-scaling are listed as follows.
\begin{itemize}
\item 
HPA \cite{zc-HPA} is a fundamental horizontal-scaling strategy in k8s framework, with the target of re-allocating resources for the dynamic workload to satisfy its demand.HPA can respond to the increasing workload by running more Pods to support overloaded traffic. On the contrary, due to the decreasing workload, HPA releases its Pods to the configured minimum.
\item 
AWS Lambda function scaling\cite{zc-Lambda} supports a commercial scaling method in the service of serverless function. Lambda can invoke a scaling strategy to avoid an overloaded service supply when the incoming traffic increases.
\item 
Knative Pod Autoscaler(KPA)\cite{zc-KPA} is an auto-scaling method supported in the recently popular framework \textit{Knative}.
KPA offers the automated scaling of applications to fit incoming demand, even for the clusters.
\end{itemize}

\subsection{Challenges and Research Opportunities}\label{sec:op_challenges}
In cloud-native environments, load balancing, service migration, and auto-scaling are essential. Load balancing optimizes resource utilization, prevents congestion, and manages workloads efficiently by considering factors such as resource allocation granularity, migration time, workload detection, and algorithm efficiency. Service migration in edge-cloud environments focuses on improving QoS, ensuring network connection continuity, and overall efficiency. Auto-scaling in cloud-native platforms involves determining the optimal monitoring interval, selecting appropriate metrics for scaling decisions, and making accurate and efficient decisions based on system states and workload predictions.
We summarize the main challenges in these three key problems as follows.
\begin{itemize}
\item \textit{Heterogeneous workloads.} Different workload has different resource demand for computing and bandwidth. The resource allocation granularity is a key for the performance of load balancing. Allocating too many resources leads to a waste while allocating too few resources causes congestion.
\item \textit{Congestion detection}. Developing efficacious algorithms to predict the unknown workload is a vital issue in cloud-native. Efficient load detection can avoid network resource congestion, especially in a resource-constraint environment.
\item \textit{Configuration management}. Migrating services often involves configuring multiple components (e.g., databases, web servers) to work together seamlessly. Keeping track of configurations and ensuring they are properly migrated can be challenging.
\end{itemize}


\section{Service Maintenance}\label{s7}
\begin{figure}[htbp]
  \centering
  \includegraphics[width=2.5in]{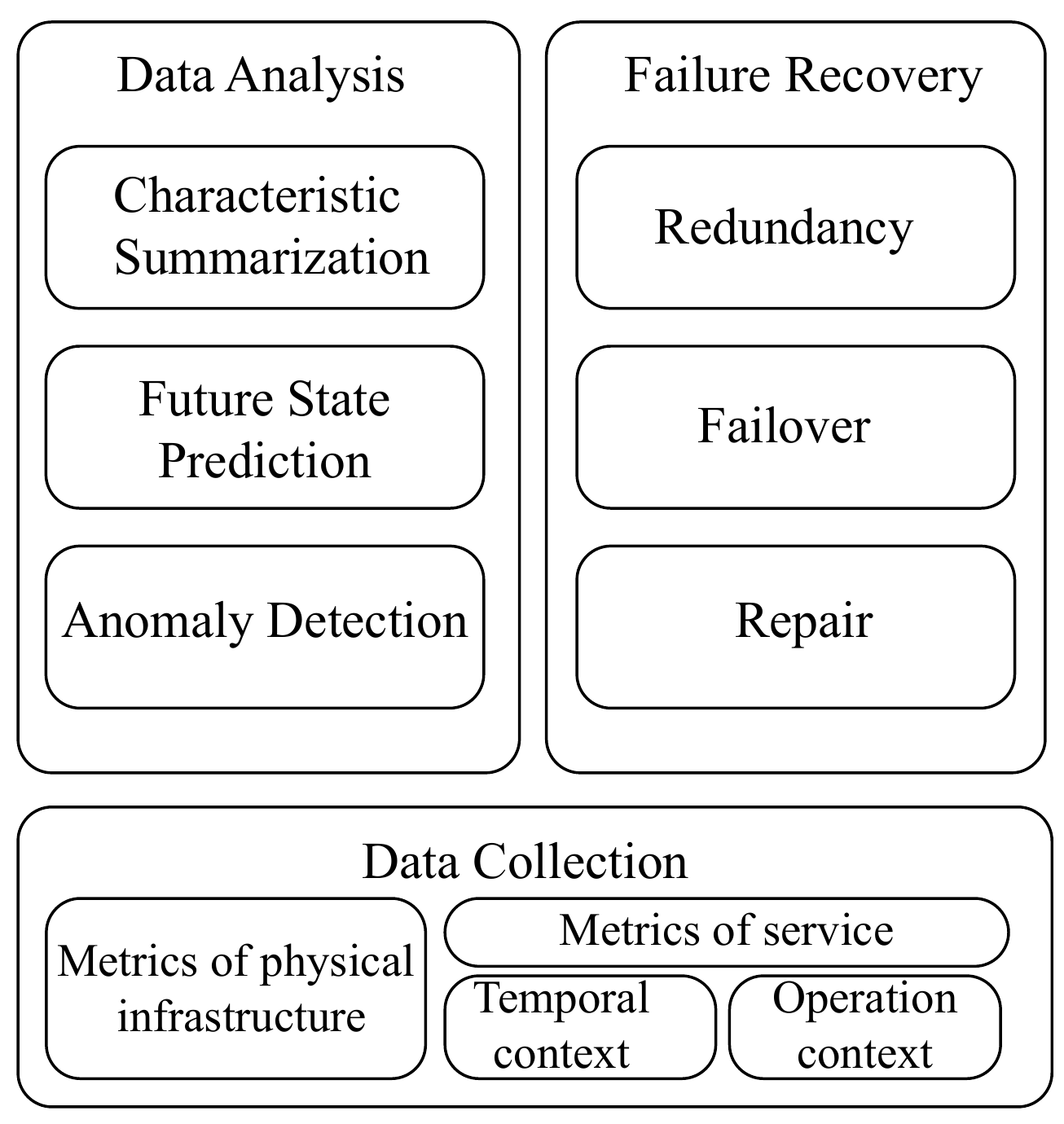}
  \caption{Organization of service maintenance.}
  \label{SermainArch}
\end{figure}

Service maintenance collects and analyzes service and system indicators, adjusts and develop strategies, and performs fault recovery.
This section is organized as Fig. \ref{SermainArch} shows. After introducing the data collection, we describe the research status of data analysis of cloud and the evolution based on data analysis. We summarize and classify the representative works based on the main purpose of them, as shown in Table \ref{represWrk}.

\begin{table*}[htbp]
\vspace{-0.15cm}
\begin{center}
    \caption{\label{represWrk}Representative works in service maintenance.}
    \begin{tabular}{c|c|c|c|c}
    \toprule
    \textsc{Work}                                                         & \textsc{QoS}        & \textsc{Server Utilization} & \textsc{Risk Control} & \textsc{Cost and Profit} \\
    \midrule
    \cite{birke2014failure,ford2010availability,padmanabhan2006study}           & \checkmark &                    & \checkmark   &                 \\
    \cite{lu2017imbalance}                                                      &            & \checkmark         &              & \checkmark      \\
    \cite{jiang2020burstable}                                                   & \checkmark & \checkmark         &              & \checkmark      \\
    \cite{ardelean2018performance}                                              & \checkmark & \checkmark         &              &                 \\
    \cite{zhu2019novel,8902077}                                                 & \checkmark & \checkmark         &              &                 \\
    \cite{7426410,de2020deep,8806998,10.1145/3225058.3225106,zeng2020detection} & \checkmark &                    & \checkmark   &                 \\
    \cite{zhang2019cross,pimentel2020deep}                                      &            &                    & \checkmark   & \checkmark      \\
    \cite{ma2021jump}                                                           & \checkmark &                    & \checkmark   &                 \\
    \cite{peng2021lime,8034965}                                                 &            &                    & \checkmark   &                 \\
    \cite{li2021fighting,wang2018cloudranger,demirbaga2021autodiagn}            &            &                    & \checkmark   & \checkmark   
    \\
    \bottomrule
    \end{tabular}%
\end{center}
\vspace{-0.15cm}
\end{table*}

\subsection{Data Collection}
Data collection aims to collect metrics of cloud service and physical infrastructure to guide strategy development.
There are mainly two aspects to collect and analyze the performance of cloud service: firstly, monitor the whole physical infrastructure of cloud environment. Secondly, monitor the performance of each service. 
For the former one, the healthy state of disks, the usage of memory, processors, network bandwidth are usually focused on. There are lots of mature tools to detect the metrics of cloud infrastructures: such as DX \cite{DX-cfy}, CA \cite{CA-cfy}, etc.
When analysing the performance of cloud services, there are mainly two kinds of contexts: the temporal context and the operation context \cite{ardelean2018performance}. The operation context means the metrics directly related to the service at each level, from software to the kernel. For example, the cache, processors, memory usage directly caused by the monitoring service. Besides, the temporal context is also important. The temporal context is the behavior of the services which have resource competition relationship with monitoring service. The challenge for collecting the operation context is that it is difficult to trace the the operation from layer to layer (i.e. from service layer to kernel layer). DTrace can solve this problem by instrumenting all code \cite{dtrace-cfy}. Besides, Ardelean et al. propagate the operation by system call \emph{getid}. \emph{getid} will ignore all the arguments passed to it, but the kernel trace will record the arguments. Thus, they use the arguments to inject operation information. As for temporal context, Magpie \cite{barham2004using-cfy} collects all the requests across multiple nodes. But it is not suitable for the cloud environment with billions users. To reduce the overhead of Magpie, a common way is to use bursty tracing \cite{arnold2002online}, which just samples partial temporal context. 
\subsection{Data Analysis}
Data analysis at service maintenance state is to mine the the property and regular patterns of running jobs from monitoring indexes and guide jobs allocation and scaling. Generally, there are three main research directions for data analysis: analyze and summarize the characteristic of running jobs and anomalous events to provide better understanding of cloud services, predict the future state of the cloud services, detect cloud service anomaly and find root cause automatically.
We organize this section as following: in section \ref{DA_Cha}, we try to answer the two questions: what kinds of data is worth and suitable to analyze, what kinds of analysis are useful for cloud providers?
 In section \ref{DA_Pred}, we introduce the research status of system state prediction, which includes the prediction of workload, prediction of healthy state of disk, prediction of service failures. In section \ref{DA_ADet}, we introduce the main challenges of anomaly detection for cloud services and the research states for each challenges.
\subsubsection{Summarize Cloud Services Characteristic}
\label{DA_Cha}
It is important to summarize the characteristic of cloud systems and the running cloud jobs, as it provides valuable insight to improve utilization of servers and quality of services. There are two important questions to summarize the characteristics. Firstly, what kind of data would contain valuable information and suitable to be analyzed? Secondly, researchers and cloud service providers are interested on what kind of characteristics? For the first question, Hauswirth et al. claimed the virtualization introduced by cloud native environment provides a significant challenge to understanding complete system performance, not found in traditionally compiled languages \cite{10.1145/1035292.1028998}. Thus, they proposes vertical profiling to provide profiling of all level of the execution stack. Vertical profiling can just apply to java-based applications. To make it more general, Ardelean et al. extend it to application based on any language \cite{ardelean2018performance}. Besides, code snippets \cite{10.1145/3503221.3508411}, functional-level variance \cite{su2019pinpointing}, control flow \cite{laguna2014diagnosis} are also suitable data. For the second question, characteristic summarizations usually concentrate on reducing the cost of providers, improving the quality of services and the healthy state of cloud systems. We list some of the most popular topics below: 
the different and similar impacts of different failures \cite{birke2014failure}, the causes of failures \cite{ford2010availability,padmanabhan2006study}, the causes of low utilizations \cite{lu2017imbalance}, pricing strategy analysis \cite{jiang2020burstable}, the analysis of time-varying mixture load \cite{ardelean2018performance}. 
\subsubsection{Future State Prediction}
\label{DA_Pred}
The cloud system and service future state prediction on the one hand can alarm forthcoming failures and risks so as to prevent them. On the other hand, it can provide basic information for task scheduling, auto-scaling, service migration and so on to improve the utilization of cloud infrastructures.
The prediction of future system state focuses on three aspects: the workload, the healthy state of cloud servers, the forthcoming service failures. 

\textbf{Workload prediction.} 
Workload prediction focus on predicting the future processor usage, memory usage and bandwidth usage, which can help to improve the quality of service(Qos) as well as improve the utilization of cloud servers.
There are mainly two kinds of workload prediction methods: the statistical methods and neural-network based methods. For statistical methods, AR \cite{1566594}, MA \cite{1566594}, ARIMA \cite{6881647}, Bayesian models \cite{SHYAM2016144} are used to predict the future workloads. Among them, ARIMA is one of the most popular and classical methods, which assumes the value at present is affected by the trend information, the history values and some noises.  
One of the problem of using ARIMA is that the workload of different jobs can have different regular patterns and it will lead to low accuracy to predict the workload by a single model. Thus, an adaptive statistic model \cite{singh2019tasm} is proposed to solve this problem, which combines linear regression, ARIMA, and support vector regression. 

Recently, it is reported that these statistic methods rely on strong mathematical assumptions (e.g. ARIMA is based on the assumption that the time series should be stationary after difference), and predict inaccurately when the workload is highly variable \cite{8902077}. Thus, many researchers turn to neural-network based methods such as RNN \cite{8281183}, LSTM \cite{KUMAR2018676}, etc. LSTM is one of the classical time series prediction methods, which can capture both the long-term dependent information and the short-term dependent information. But these recurrent network based methods give the same weights to the workload in observing window, while the history workload has different impact on predicting workload. Thus, a method combines LSTM and attention mechanism is proposed to put different weight on history workload \cite{zhu2019novel}. Besides, another problem of using recurrent network based methods is the forgetting effect \cite{zhou2021informer} when extracting long-term dependent information. Thus, a method \cite{8902077} combines top-sparse auto-encoder and GRU is proposed to effectively extract the essential representations of workloads from the original high-dimensional workload data and predict highly variable workload accurately.

\textbf{Healthy state of cloud servers prediction.} 
Researchers in this domain mainly focus on disk dives failure prediction, as it can dramatically reduce data restoring time to predict disk failures in advance.
Disk drives failure prediction plays a very important and crucial role in reducing data center downtime and significantly improving service reliability \cite{de2020deep}, as it alarms forthcoming disk drives failure and the system can overlap the time of regular data operation and the time of data restoring. At the beginning, the task of disk drives failure prediction is regarded as a binary classification problem and lots of classification models are used to predict the disk failure, such as Bayesian models \cite{hamerly2001bayesian}, Wilcoxon rank-sum test \cite{hughes2002improved}, support vector machines (SVM) \cite{murray2005machine} artificial neural network (ANN) \cite{russell1995artificial}, etc. 

However, these methods have reported poor performance on real-world environment. Firstly, the status of disk drives corrupt gradually and is not only either good or bad \cite{7426410} \cite{de2020deep}. Thus, Aniello et al. propose a method firstly divide the healthy status of disks drives into seven levels by regression tree, and then use LSTM to predict the disk healthy status in the future \cite{de2020deep}. The fine-grained disk drivers healthy status support more flexible data-restore mechanism, which can plan data restoring in advance according to the different prediction of fine-grained disk drivers healthy status. However, LSTM used in this work is recurrent networks and it is reported to be vulnerable to he highly variant interval between triggering events and hardware failures \cite{8806998}.
 Thus, Sun et al. use the temporal Convolution Neural Network (CNN) to leverage CNN’s characteristic of translation invariance, which can make the CNN insensitive to various delays between triggering-and failure-events in the time dimension \cite{8806998}. Secondly, the data imbalance between disk failure data and normal data hinders the models to predict accurately. Thus, Sun et al. also design a new loss function to prevent the gradient vanishing in front of the huge data imbalance \cite{8806998}. Moreover, the above methods are based on offline training and can not adapt to the continuous update systems \cite{10.1145/3225058.3225106}. Thus, Xiao et al. proposes a method based on online random forest algorithm to maintain stable predicting accuracy for long-term usage.

\textbf{Service failure prediction.}
 Different from the above sections, service failure prediction focus on service Qos and predict the failure from the level of service. The service failure can lead to penalty payments, profit margin reduction, reputation degradation, customer churn and service interruptions \cite{zeng2020detection}.
Thus, it is worthwhile to know the possible failure in advance. By doing so the cloud systems can take steps to prevent the predicting failures. 

Generally, service failure prediction can be divided to three categories: rule based methods, statistic methods, deep learning methods. Rule based methods rely on manually defined rules to predict the failures, which are limited to the human experience and its adaptability is poor. The other two methods are data-driven method, compared with rule based methods, they are more flexible and convenient.

The rule based methods require experts define specific rules in advance. For example, PerfAugur \cite{roy2015perfaugur} is designed to predict failures by specified features. Generally, these methods are accurate but just suitable for specific scenarios. 

The statistic methods include ARMA \cite{chalermarrewong2012failure}, ARIMA \cite{godse2010automating}, SVM \cite{fronza2013failure}, Hidden Semi-markov Model \cite{fu2007exploring}, etc. Among the classical statistic methods, Cavallo et al. \cite{cavallo2010empirical} have claimed that ARIMA forecasting has the best compromise in ensuring a good prediction error, being sensible to outliers, and being able to predict likely violations of QoS constraints \cite{6257792}. However, Amin et al. \cite{6257792} point out the traditional ARIMA model can not deal with the high volatility of quality of service (QoS) properly. Thus, they propose a model integrate GARCH and ARIMA to solve this problem. 

However, statistic methods rely on some mathematics assumptions and can not work well on high dimension feature and dependent sequence data. Thus, many researchers apply the deep learning models to service failure prediction. Chen et al. \cite{chen2014failure} propose a deep learning methods based recurrent neural networks (RNN) to predict task-level failures. However, the drawback of RNN is that it will definitely forget the information in long distance, which will degrade the predicting accuracy. Although some modified recurrent neural networks, such as LSTM \cite{du2017deeplog}, can mitigate this effect, the weights put on each value in observing window is unequal and degrade as the distance goes farther. Thus, Gao et al. \cite{9090992} propose a method based on bi-direction LSTM to further improve the accuracy.

\subsubsection{Anomaly Detection}
\label{DA_ADet}
\begin{figure}[htbp]
  \centering
  \includegraphics[width=3.3in]{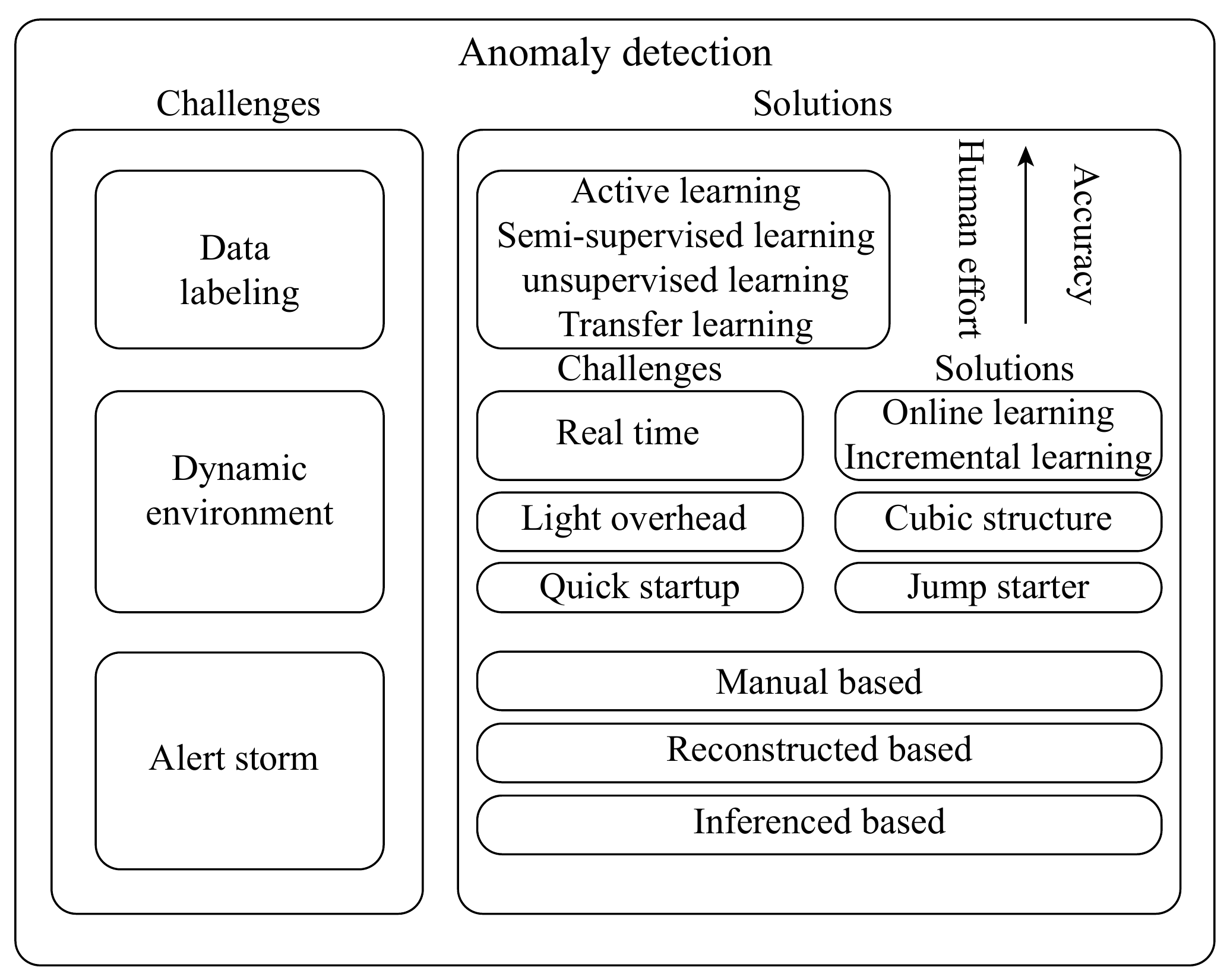}
  \caption{Organization of anomaly detection.}
  \label{AnomaDetct}
\end{figure}

In the cloud native scene, researchers detect anomaly from different aspects: components anomalous usage (e.g. anomalous processor and memory usage, network attacks, disk drives failures), service anomalous Qos (e.g. high latency and low throughput). It is worth to notice that though anomaly detection and system state prediction both study different component usages and Qos, the prediction is to infer the forthcoming system states, while anomaly detection is to discover the anomaly already happens. The organization of this subsection is illustrated in Fig.\ref{AnomaDetct}. There are three main challenges when detecting anomaly: labeled data obtaining issues, high variance of cloud environment, alert storm. The labeled data obtaining issues is caused by the requirements of expertise experience and lots of efforts to label the anomalous data. The high variance of cloud environment suggests that the detecting patterns the models learned from history data may become outdated frequently. Besides, there are lots of APIs and components in cloud environment. Lots of them are relevant to each other. When anomaly occurs to one part, the related parts will also be abnormal. Thus, when an anomaly occurs, the system always suffers from a alerting storm. This phenomena suggests the necessity of root cause finding.

\textbf{Data labeling issue.}
The anomalous data labeling in cloud is different from labeling tasks in many other domains (e.g. image recognition), as it requires expertise experience to label the anomaly, which makes the labeled data rarer.
There are mainly four kinds of methods to solve it: semi-supervised learning \cite{mohseni2020self,sun2022gradient,daneshpazhouh2014entropy,erfani2016high}, unsupervised learning \cite{zhang2019deep,chen2022deep}, transfer learning \cite{zhang2019cross,wen2019time,michau2021unsupervised}, active learning \cite{zhang2019cross,pimentel2020deep}. 
Semi-supervised learning is suitable for the datasets with small amount of labeled data. It firstly uses the labeled data to train a primary model. After that, it uses the primary model to assign pseudo label to the remaining unlabeled data. Then, it uses the data with pseudo label and data with label to retrain the primary model. If we not only have a dataset with small amount of labeled data, but also have experts to label some data in the process of model training, we can use the active learning to further improve the detecting accuracy. Active learning firstly uses the labeled data to train a primary model. After that, it uses the primary model to pick a small part of unlabeled data which need label most and label the remaining data with pseudo labels. Then, experts label the data picked by the primary model. Then, it retrains the primary model with newly labeled data, labeled data, pseudo labeled data. However, if we have none labeled data at all, we can use transfer learning or unsupervised learning. Both of them rely on assumptions: transfer learning assume there are some similarities between source dataset and target dataset, unsupervised learning assume the normal state has some unified and invariant latent regularity that can be learned by models. Autoencoder \cite{borghesi2019anomaly} is one of the most classical unsupervised learning models in anomaly detection. It firstly compresses the features into a smaller latent variable. After that, it reconstructs the features from the latent variable. In training process, it makes the reconstructed features as similar as possible to the original features. It assumes in the training process, autoencoder can learn the reconstructing patterns of normal data and when the autoencoder meets anomalous data in inferring process, it will fail to reconstruct. By then, autoencoder can detect anomalies. As each feature in cloud anomaly detection is time dependent, recently, some researchers combine the recurrent neural network with the autoencoder to capture the correlation between features as well as the time dependence \cite{zuo2020intelligent,shen2021time}. 

Generally speaking, the detecting accuracy of active learning, semi-supervised learning, unsupervised learning and transfer learning decreases one by one, as well as the inputting human efforts. Therefore, the choice of model depends on the constraints and main objectives in the practical environment.

\textbf{Cloud environment high variance issue.}
The cloud environment is highly variable. For example, Google and baidu are reported thousands of software changes are deployed everyday \cite{ma2021jump}. These changes will continuously degrade the predicting accuracy of anomaly detection models trained by outdated data. Frequent model retraining is costly and impractical. This problem calls for lightweight and self-evolving algorithms. The data-driven anomaly detection methods on the cloud scene can be roughly divided to two streams: one is based on time dependent information, the other is based on the correlation between different metrics. The former one is to capture the time-dependent normal pattern from time series and compare the tested data with the normal pattern, while the latter one is to capture the normal correlation of different metrics and compare the correlation of tested metrics with the normal correlation. For the former anomaly detection methods, Xu et al. \cite{8034965} propose a method based on online machine learning to mitigate the negative effect of cloud environment variance. Besides, they also propose a reduction method to further reduce the overhead of retraining and inferring. For the latter anomaly detection methods, Peng et al. \cite{peng2021lime} propose a method uses cuboid structure to store the relationship of different metrics, which significantly reduce the storing overhead. Besides, they also propose a incremental learning method suitable for cuboid structure, which significantly reduce the retraining overhead. However, it is reported the incremental retraining methods need lots of data points to converge \cite{xu2018unsupervised} and still need more data to reach steady state \cite{su2019robust}, which requires the cloud system collect enough data points before retraining the models. The data collecting time is long (generally tens of days) and there will be a period called initial time \cite{ma2021jump} when the accuracy of old model is low and there is no enough data to apply the incremental learning. To reduce the initial time, Ma et al. \cite{ma2021jump} propose a quickly start method which can work well without relying on as big amount of data as incremental learning and report high accuracy.

\textbf{Alert storm issue.}
Cloud system is consisted of thousands of components with extraordinary complex dependency \cite{li2021fighting}. Besides, business transactions in a cloud native system usually have a much longer calling path with dozens of distributed microservices participating \cite{wang2018cloudranger}. Thus, an anomaly in one part will trigger lots of anomaly alerts in related components and tasks calling the anomalous task, which is called Alert storm \cite{wang2018cloudranger}. This problem calls for root cause finding techniques. 
There are mainly three kinds of root cause finding techniques: manual-rule based methods, causal inference based methods, reconstruction based methods. Manual-rule based methods are based on expertise experience, which is accurate though human-work costly and easily-outdated. For example, Demirbaga et al. \cite{demirbaga2021autodiagn} defined three kinds anomaly cause: data locality (i.e. data needed is not at the same server as task), resource heterogeneity (i.e. jobs are scheduled to machine remaining few computing resource), network failure (i.e. network disconnection). They detect these anomaly by defining threshold of several metrics, when the defining combined metrics exceed the thresholds, then corresponding anomaly is detected. Due to the limitations of manual-rule based methods, researchers also pay attention to data-driven methods. Some of them are just suitable for specific anomaly detection methods, such as reconstruction based methods \cite{zhang2019deep}. They can just apply to the anomaly detecting methods based features reconstruction. This root cause finding methods work by compute the distances between every pair of reconstructed feature and original feature. The greater the distance is, the more the feature contributes to the anomaly. More generally, causal inference based methods can apply to more kinds of anomaly detection methods, though they are more computationally expensive. CloudRanger \cite{wang2018cloudranger} use conditional dependence to establish the topology of cause and effect relationship among tasks and use the topology to find root cause. Sipple \cite{sipple2020interpretable} uses integrated gradient method to compute the contribution of each feature and find the root cause.

\subsection{Failure Recovery}
One of the main advantages of cloud-native solutions lies in the possibility of automatically detecting and overcoming failures. Failure recovery is made possible by multiple technologies leveraged in building a cloud-native application. First, containerization technologies like Docker introduce almost zero overhead when launching a container, making it possible to re-spawn a failed container within seconds possible. Second, modern metric collection and processing systems like Prometheus\cite{dyn-Prometheus} allow high integration into cloud-native systems, further enabling well-informed decisions.
Third, with fast development of machine learning techniques, automatically-made decisions are getting yet more efficient and effective. In this section, we will study three main topics to recover from an error: redundancy, failover, and repair. 

\subsubsection{Redundancy} 
Redundancy refers to deploying multiple replicas of one resource (microservice, computation, storage, etc.), preferably in multiple locations, in order to minimize disruption even if one or several of them goes down. This method has already been in use before the container-based cloud-native era \cite{dyn-abu2010racs}. In addition to improved reliability, distributing data from multiple locations can lead to reduced latency to end users. Kang et al. \cite{dyn-kang2021implementation} design and implement a custom controller in Kubernetes to select and use multiple replicas of Virtual Network Functions (VNFs) so high processing ability can be achieved. However, making $N$ redundant copies means $N$-time of resource consumption, which could be costly. Uluyol et al. \cite{dyn-uluyol2020near} propose a novel encoding mechanism to save encode data before saving to multiple locations, instead of simply creating replicas. Furthermore, the authors mitigated increase in latency introduced by distributed storage by rethinking how consensus can be reached, to offer near-optimal latency versus cost trade-offs.

\subsubsection{Failover}
Failover is the process of switching to a backup server or other types of resource when disruption is detected. This process is based on redundant deployment talked in the previous section. By having proper algorithms configured, the cloud orchestrator is able to make efficient use of replicas to switch to other healthy replicas. There are multiple optimization targets in the context of failover. Aldwyan et al. \cite{dyn-aldwyan2019latency} identify that failover between distributed data centers can lead to degraded performance due to added network latencies, and propose a latency-aware failover strategy leveraging genetic algorithms to take latancies into consideration when making a failover decision. Jin et al. \cite{dyn-jin2019fave} build a SDN failover mechanism, FAVE, that is aware of physical link failure, to be used in virtualized SDN environments. Landa et al. \cite{dyn-landa2021staying} utilize TCP re-transmission metrics to declare network failure in CDN networks, and quick re-route traffic through redundant links to keep high availability. 

\subsubsection{Repair}
Repair tries to fix the error instead of redirect traffic to other service instances. Considering the nature of today's container-based cloud-native solutions, repairing a failed service is likely to be more efficient compared to failover into backup. Giannakopoulos et al. \cite{dyn-giannakopoulos2018cloud} consider the complexity in modern cloud deployments and identify that such complexity could lead to failure in deployments. They build AURA, which transform a deployment into a directed acyclic graph, so whenever an error happens, it is possible to respawn only a small portion of the entire deployment, thus keeping the repair process efficient.

\subsection{Challenges and Research Opportunities}
\textit{The commonality and special individuality of data distribution.} In data-driven models training, either for future state prediction or for anomaly detection, there is a main concern that whether to train only one model for all the servers or train single model for each server.
On the one hand, it is cost to train a model for each server.
On the other hand, it predicts inaccurately when training only a model for all the servers, as every server has its own data distribution. Thus, there is a trade off between efficiency and accuracy.

\textit{The dynamic evolution of cloud environment.} The cloud environment is dynamic. New missions arrives and old missions ends at every moment. The models generally becomes outdated and need retraining frequently. Designing a light-weight retraining methods can bring huge benefit.


\section{Open Issues and Future Directions}\label{s8}
Cloud-native computing has been gaining a lot of attention in recent years due to its ability to enable agile, scalable and resilient software systems. However, there are still some open issues and future directions that need to be addressed. In the following, we list the open issues and possible research directions.

\begin{itemize}
  \item \textit{Hybrid multi-cloud integration}. As the popularity of cloud-native computing continues to grow, many organizations are using multiple cloud providers or leveraging both public and private clouds. Besides, the services are applicaitons are deployed across the continuum of cloud-edge-device. It is important to develop better tools and techniques with interoperability capabilities for integrating these environments, including standardization of APIs and data exchange while retaining control over sensitive private data across multiple clouds.
  \item \textit{User-friendly servie shapes (forms).} As cloud-native being the foundation of today's most web applicaitons, more user-friendly service shapes to erase the heavy burden of application deployment for hybrid edge-clouds are urgently needed. Serverless computing is a good practice. It allows developers to focus on their business logic without worrying about the infrastructure management. However, serverless computing is criticized by its long cold time, inefficient state management and other related issues. Better service shapes/forms for more wider application scenarios are required.
  \item \textit{Advanced automation and resource utilization}. Automation is crucial to realizing the full benefits of cloud-native architectures. However, there is still a lot of room for improvement in terms of automating deployment, scaling, and maintenance activities, especially for the distributed training of the heavy big models. Another key benefit of cloud-native computing is the ability to dynamically allocate resources based on demand. However, this can also lead to inefficiencies if not properly managed. There is a need for improved tools and algorithms that can optimize resource allocation and reduce waste.
  \item \textit{Enhanced cross-platform observability and security}. Cloud-native architectures tend to be highly distributed and dynamic, which can make it difficult to observe and troubleshoot issues. There is a need for better observability and monitoring tools to help developers and operations teams quickly identify and resolve problems. Security related concerns are becoming more critical, especially for hybrid multi-cloud scenarios. There is a string need for better security mechanisms that can effectively protect against cyber threats, especially as attacks become more sophisticated.
\end{itemize}
In summary, while cloud-native computing has come a long way, there are still many open issues and future directions that need to be addressed to fully realize its potential. By continuing to innovate and address these challenges, we can create more efficient, secure, and scalable software systems for the future.


\section{Conclusions}\label{s9}
Cloud-native, as the most influential principle for web applications, has attracted more and more researchers and companies to get involved in studying and using it. This survey attempts to provide possible research opportunities through a succinct and effective classification. We present the research roadmap of cloud-native from the perspective of services computing. Specifically, we divide the development of cloud-native applications into four states, building, orchestration, operate, and maintenance. State-of-the-art research works and industrial applications are provided. We attempted to provide some enlightening thoughts on the research of cloud-native computing and services computing. We hope that this article can stimulate fruitful discussions on potential future research directions on this topic.

\ifCLASSOPTIONcompsoc
  \section*{Acknowledgments}
\else
  \section*{Acknowledgment}
\fi
This work was supported in part by the National Key Research and Development Program of China under Grant 2022YFB4500100, the National Science Foundation of China under Grants 62125206 and U20A20173, and the Key Research Project of Zhejiang Province under Grant 2022C01145.

\bibliographystyle{IEEEtran}
\bibliography{IEEEabrv,global_ref.bib}


\begin{IEEEbiography}
  [{\includegraphics[width=1in,height=1.25in,clip,keepaspectratio]{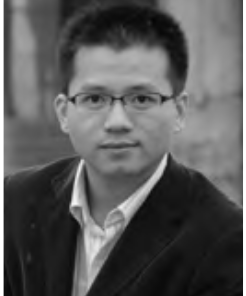}}]{Shuiguang Deng} 
  is currently a full professor at the College of Computer Science and Te  chnology in Zhejiang University, China, 
  where he received a BS and PhD degree both in Computer Science in 2002 and 2007, respectively. He previously 
  worked at the Massachusetts Institute of Technology in 2014 and Stanford University in 2015 as a visiting scholar. 
  His research interests include Edge Computing, Service Computing, Cloud Computing, and Business Process Management. 
  He serves for the journal IEEE Trans. on Services Computing, Knowledge and Information Systems, Computing, and IET 
  Cyber-Physical Systems: Theory \& Applications as an Associate Editor. Up to now, he has published more than 100 
  papers in journals and refereed conferences. In 2018, he was granted the Rising Star Award by IEEE TCSVC. He is 
  a fellow of IET and a senior member of IEEE.
\end{IEEEbiography}

\begin{IEEEbiography}
  [{\includegraphics[width=1in,height=1.25in,clip,keepaspectratio]{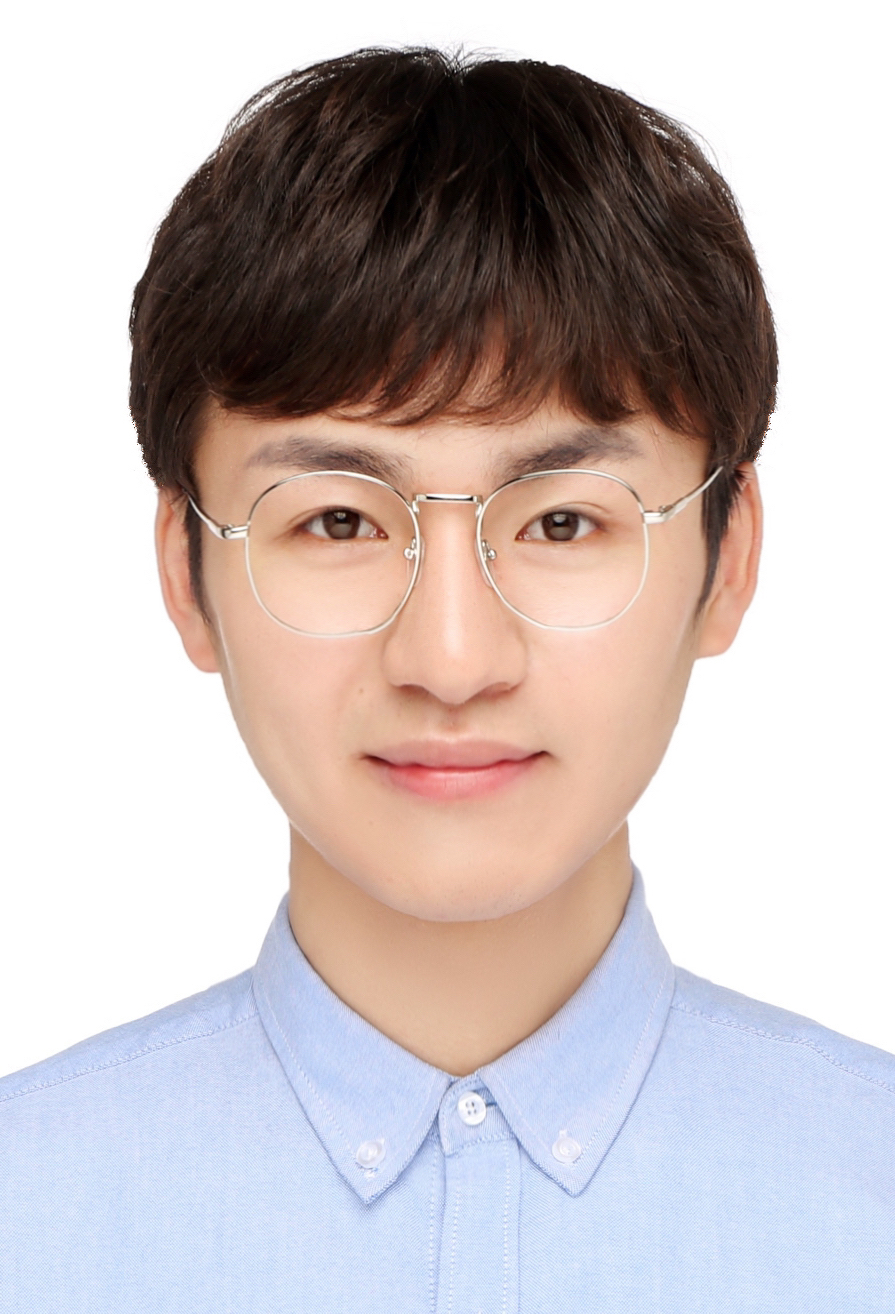}}]{Hailiang Zhao} 
  received the B.S. degree in 2019 from the school of computer science and technology, Wuhan University of 
  Technology, Wuhan, China. He is currently pursuing the Ph.D. degree with the College of Computer Science 
  and Technology, Zhejiang University, Hangzhou, China. His research interests include cloud \& edge computing, 
  distributed computing and optimization algorithms. He has published several papers in flagship conferences 
  and journals including IEEE ICWS 2019, IEEE TPDS, IEEE TMC, etc. He has been a recipient of the Best 
  Student Paper Award of IEEE ICWS 2019. He is a reviewer for IEEE TSC and Internet of Things Journal.
\end{IEEEbiography}

\begin{IEEEbiography}
  [{\includegraphics[width=1in,height=1.25in,clip,keepaspectratio]{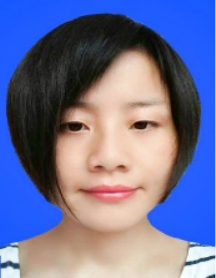}}]{Binbin Huang} is an 
  Assistant Professor in the College of Computer Science at the University of Hangzhou Dianzi, in Hangzhou, China. 
  She received her PhD degree in Computer Science and Technology from Beijing University of Posts and 
  Telecommunications in 2014. His research interests include cloud computing, mobile edge computing, and 
  reinforcement learning.
\end{IEEEbiography}

\begin{IEEEbiography}
  [{\includegraphics[width=1in,height=1.25in,clip,keepaspectratio]{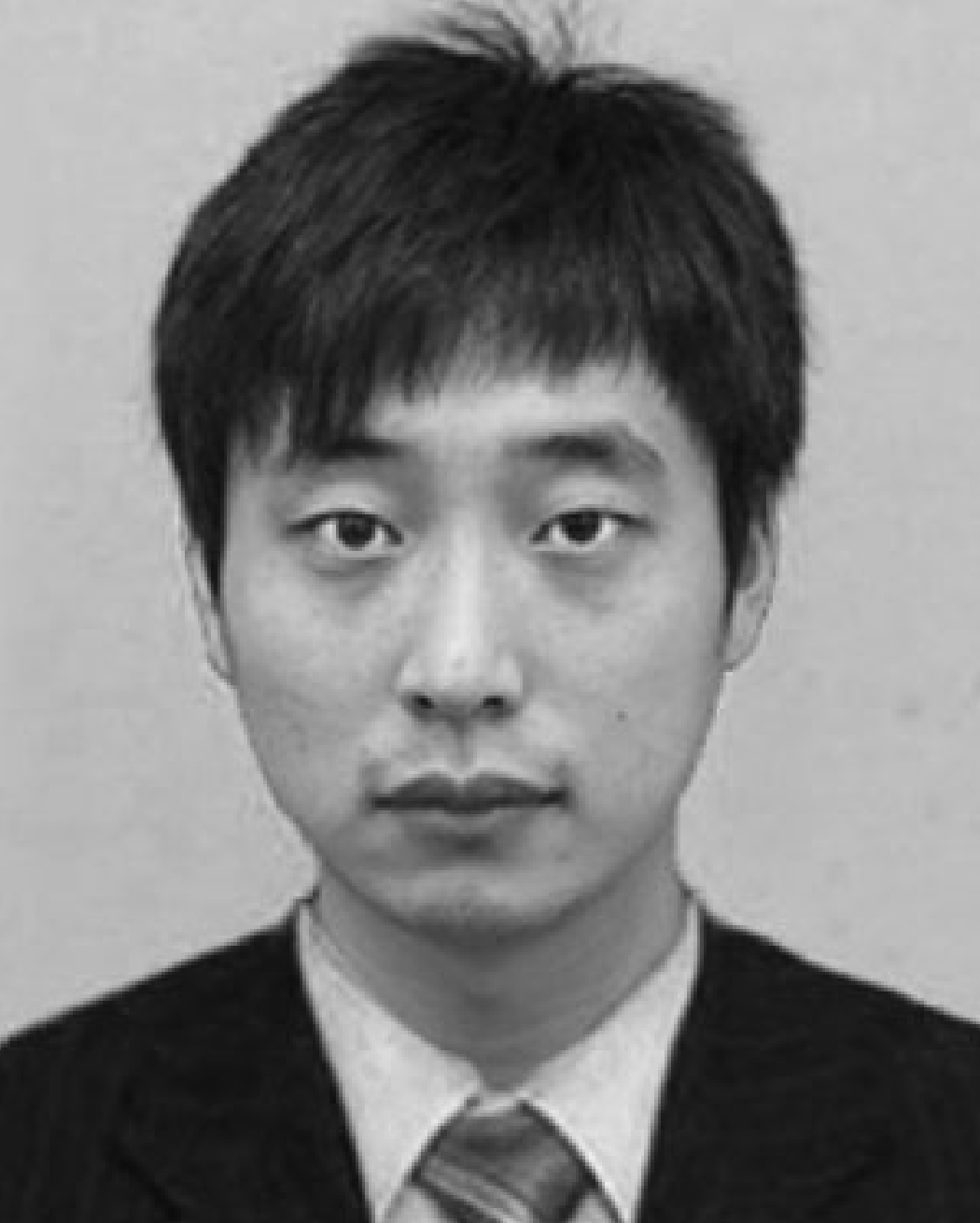}}]{Cheng Zhang}
  received the MS degree in electrical engineering from Zhejiang University, China, in 2013, Currently, he is 
  working toward the PhD degree in computer science and technology at Zhejiang University. His research interests 
  include edge computing and edge intelligence.
\end{IEEEbiography}

\begin{IEEEbiography}
  [{\includegraphics[width=1in,height=1.25in,clip,keepaspectratio]{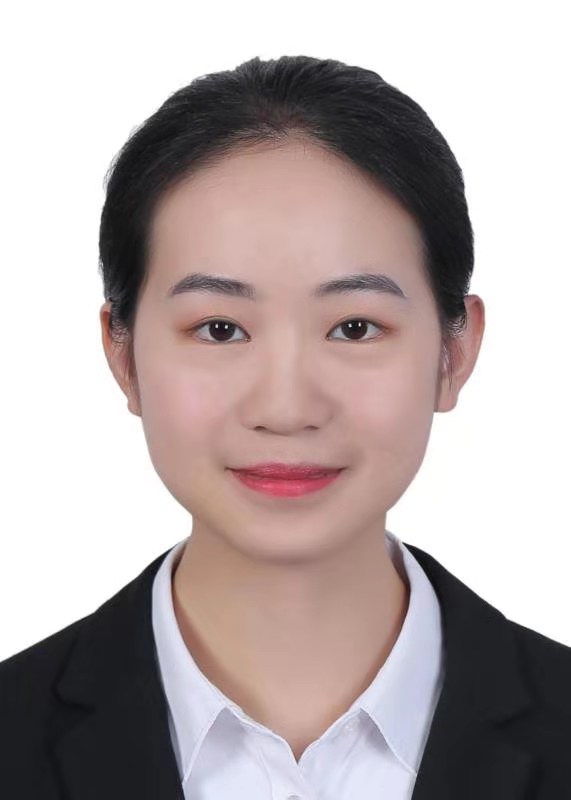}}]{Feiyi Chen} 
  received the B.S. degree in 2021 from the school of computer science and engineering, Sun Yat-sen 
  University (SYSU), Guangzhou, China. She is currently pursuing the master degree with the College of 
  Computer Science and Technology, Zhejiang University, Hangzhou, China. Her research interests include 
  cloud computing, edge computing, and distributed systems.
\end{IEEEbiography}

\begin{IEEEbiography}
  [{\includegraphics[width=1in,height=1.25in,clip,keepaspectratio]{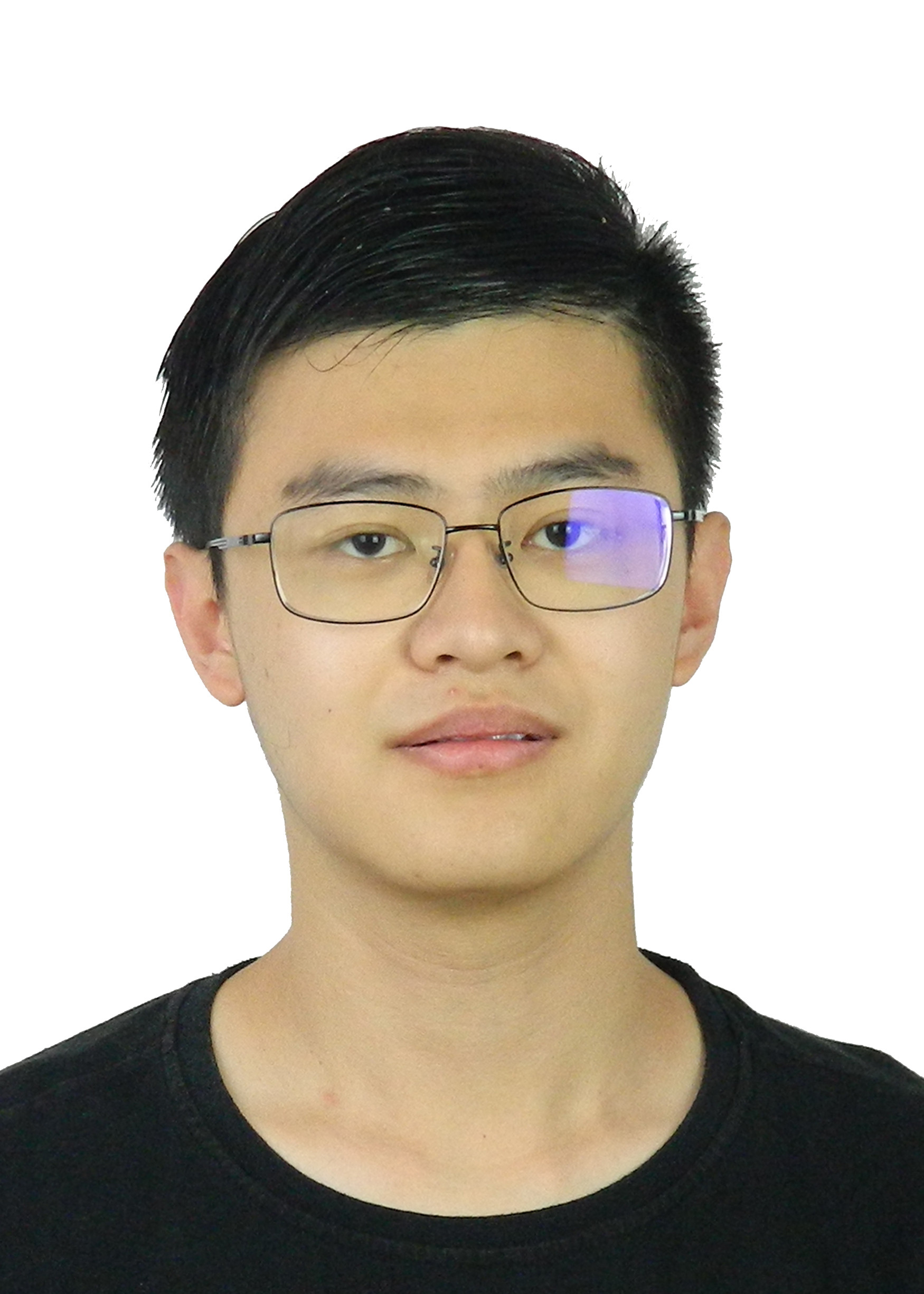}}]{Yinuo Deng} received the B.S. degree in 2022 from the School of Artificial Intelligence, Beijing University of Posts and Telecommunications, Beijing, China. He is currently a M.S. student in computer science of technology at College of Computer Science and Technology, Zhejiang University, Hangzhou, China. His research interests include cloud computing, networking, and distributed systems.
\end{IEEEbiography}

\begin{IEEEbiography}
  [{\includegraphics[width=1in,height=1.25in,clip,keepaspectratio]{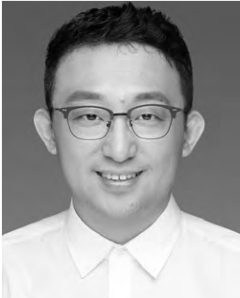}}]{Jianwei Yin} 
  received the Ph.D. degree in computer science from Zhejiang University (ZJU) in 2001. 
  He was a Visiting Scholar with the Georgia Institute of Technology. He is currently a Full Professor 
  with the College of Computer Science, ZJU. Up to now, he has published more than 100 papers in top 
  international journals and conferences. His current research interests include service computing 
  and business process management. He is an Associate Editor of the IEEE Transactions on Services 
  Computing.
\end{IEEEbiography}

\begin{IEEEbiography}
  [{\includegraphics[width=1in,height=1.25in,clip,keepaspectratio]{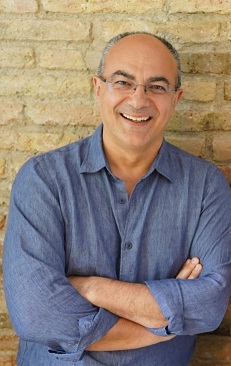}}]{Schahram Dustdar}
  is a Full Professor of Computer Science (Informatics) with a focus on Internet Technologies heading the Distributed Systems Group at the TU Wien. He was founding co-Editor-in-Chief of ACM Transactions on Internet of Things (ACM TIoT). He is Editor-in-Chief of Computing (Springer). He is an Associate Editor of IEEE Transactions on Services Computing, IEEE Transactions on Cloud Computing, ACM Computing Surveys, ACM Transactions on the Web, and ACM Transactions on Internet Technology, as well as on the editorial board of IEEE Internet Computing and IEEE Computer. Dustdar is recipient of multiple awards: TCI Distinguished Service Award (2021), IEEE TCSVC Outstanding Leadership Award (2018), IEEE TCSC Award for Excellence in Scalable Computing (2019), ACM Distinguished Scientist (2009), ACM Distinguished Speaker (2021), IBM Faculty Award (2012). He is an elected member of the Academia Europaea: The Academy of Europe, where the chairman of the Informatics Section for multiple years. He is an IEEE Fellow (2016), an Asia-Pacific Artificial Intelligence Association (AAIA) President (2021) and Fellow (2021). He is an EAI Fellow (2021) and an I2CICC Fellow (2021). He is a Member of the IEEE Computer Society Fellow Evaluating Committee (2022 and 2023).
\end{IEEEbiography}

\begin{IEEEbiography}
  [{\includegraphics[width=1in,height=1.25in,clip,keepaspectratio]{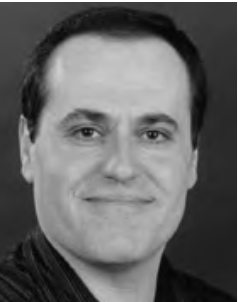}}]{Albert Y. Zomaya} is the Peter Nicol Russell Chair Professor of Computer Science and Director of the Centre for Distributed and High-Performance Computing at the University of Sydney. To date, he has published > 600 scientific papers and articles and is (co-)author/editor of > 30 books. A sought-after speaker, he has delivered > 250 keynote addresses, invited seminars, and media briefings. His research interests span several areas in parallel and distributed computing and complex systems. He is currently the Editor in Chief of the ACM Computing Surveys and processed in the past as Editor in Chief of the IEEE Transactions on Computers (2010-2014) and the IEEE Transactions on Sustainable Computing (2016-2020). Professor Zomaya is a decorated scholar with numerous accolades including Fellowship of the IEEE, the American Association for the Advancement of Science, and the Institution of Engineering and Technology (UK). Also, he is an Elected Fellow of the Royal Society of New South Wales and an Elected Foreign Member of Academia Europaea. He is the recipient of the 1997 Edgeworth David Medal from the Royal Society of New South Wales for outstanding contributions to Australian Science, the IEEE Technical Committee on Parallel Processing Outstanding Service Award (2011), IEEE Technical Committee on Scalable Computing Medal for Excellence in Scalable Computing (2011), IEEE Computer Society Technical Achievement Award (2014), ACM MSWIM Reginald A. Fessenden Award (2017), the New South Wales Premier’s Prize of Excellence in Engineering and Information and Communications Technology (2019), and the Research Innovation Award, IEEE Technical Committee on Cloud Computing (2021). 
\end{IEEEbiography}

\end{document}